%% file: main.tex
\documentclass[prb,reprint,superscriptaddress,floatfix,amsmath,amssymb]{revtex4-2}

\usepackage[titletoc,toc,title]{appendix}
\usepackage{graphicx}
\usepackage{bm}
\usepackage{xcolor}
\usepackage{booktabs}
\usepackage{hyperref}
\usepackage{diagbox}

\renewcommand{\vec}[1]{\bm{#1}}

\begin{document}

\title{Effects of Structural Distortions on the Electronic Structure of T-type Transition Metal Dichalcogenides}

\author{Daniel T. Larson}
\affiliation{Department of Physics, Harvard University, Cambridge, Massachusetts 02138, USA.}
\author{Wei Chen}
\affiliation{Department of Physics, Harvard University, Cambridge, Massachusetts 02138, USA.}
\affiliation{Center for Functional Nanomaterials, Brookhaven National Laboratory, Upton, New York 11973, USA.}
\author{Steven B. Torrisi}
\affiliation{Department of Physics, Harvard University, Cambridge, Massachusetts 02138, USA.}
\author{Jennifer Coulter}
\affiliation{John A. Paulson School of Engineering and Applied Sciences, Harvard University, Cambridge, Massachusetts 02138, USA.}
\author{Shiang Fang}
\email{shiangfang913@gmail.com }
\affiliation{Department of Physics, Harvard University, Cambridge, Massachusetts 02138, USA.}
\affiliation{Department of Physics and Astronomy, Center for Materials Theory, Rutgers University, Piscataway, NJ 08854 USA}
\author{Efthimios Kaxiras}
\email{kaxiras@physics.harvard.edu}
\affiliation{Department of Physics, Harvard University, Cambridge, Massachusetts 02138, USA.}
\affiliation{John A. Paulson School of Engineering and Applied Sciences, Harvard University, Cambridge, Massachusetts 02138, USA.}

\date{\today}

\begin{abstract}
Single-layer transition metal dichalcogenides (TMDCs) can adopt two distinct structures corresponding to different coordination of the metal atoms. TMDCs adopting the T-type structure exhibit a rich and diverse set of phenomena, including charge density waves (CDW) in a $\sqrt{13}\times\sqrt{13}$ supercell pattern in TaS$_2$ and TaSe$_2$, and a possible excitonic insulating phase in TiSe$_2$. These properties make the T-TMDCs desirable components of layered heterostructure devices. In order to predict the emergent properties of combinations of different layered materials, one needs simple and accurate models for the constituent layers which can take into account potential effects of lattice mismatch, relaxation, strain, and structural distortion. Previous studies have developed \emph{ab initio} tight-binding Hamiltonians for H-type TMDCs~\cite{fang2018electronic}. Here we extend this work to include T-type TMDCs. 
We demonstrate the capabilities of our model using three example systems: a 1-dimensional sinusoidal ripple, the 2$\times$2 CDW in TiSe$_2$, and the $\sqrt{13}\times\sqrt{13}$ CDW in TaS$_2$. Using the technique of band unfolding we compare the electronic structure of the distorted crystals to the pristine band structure and find excellent agreement with direct DFT calculations, provided the magnitude of the distortions remains in the linear regime.
\end{abstract}

\maketitle

\section{\label{sec:intro}INTRODUCTION}

There has been significant progress in the production of van der Waals heterostructures, devices formed by combinations of various two-dimensional layered materials~\cite{2dmaterial_rev1}. These layers offer a promising platform for applications in optoelectronics~\cite{2dmaterial_rev2}, spintronics~\cite{TMDC_spintronic}, valleytronics~\cite{valleytronics}, straintronics~\cite{ong2012engineered,bukharaev2018straintronics}, twistronics~\cite{Twistronics_DFT},  nanomechanical resonators~\cite{MoS2_resonator}, and plasmonics~\cite{tmdc_plasmon}.
Individual layers can be insulators, semimetals, or metals, and can express many different quantum orders, including charge density waves (CDW)~\cite{ritschel2015orbital}, superconductivity (SC)~\cite{TMDC_sc}, magnetism~\cite{TMDC_magnetism,2d_magnetism1,2d_magnetism2}, topological phases~\cite{TMDC_TI,WTe2_QSHE}, and Mott insulator physics~\cite{nakata2016monolayer,TTMDC_TaS2_Mott,TaS2_QSL,PLT_Mott}.

\begin{figure}[htbp]
\centering
\includegraphics[width=\columnwidth]{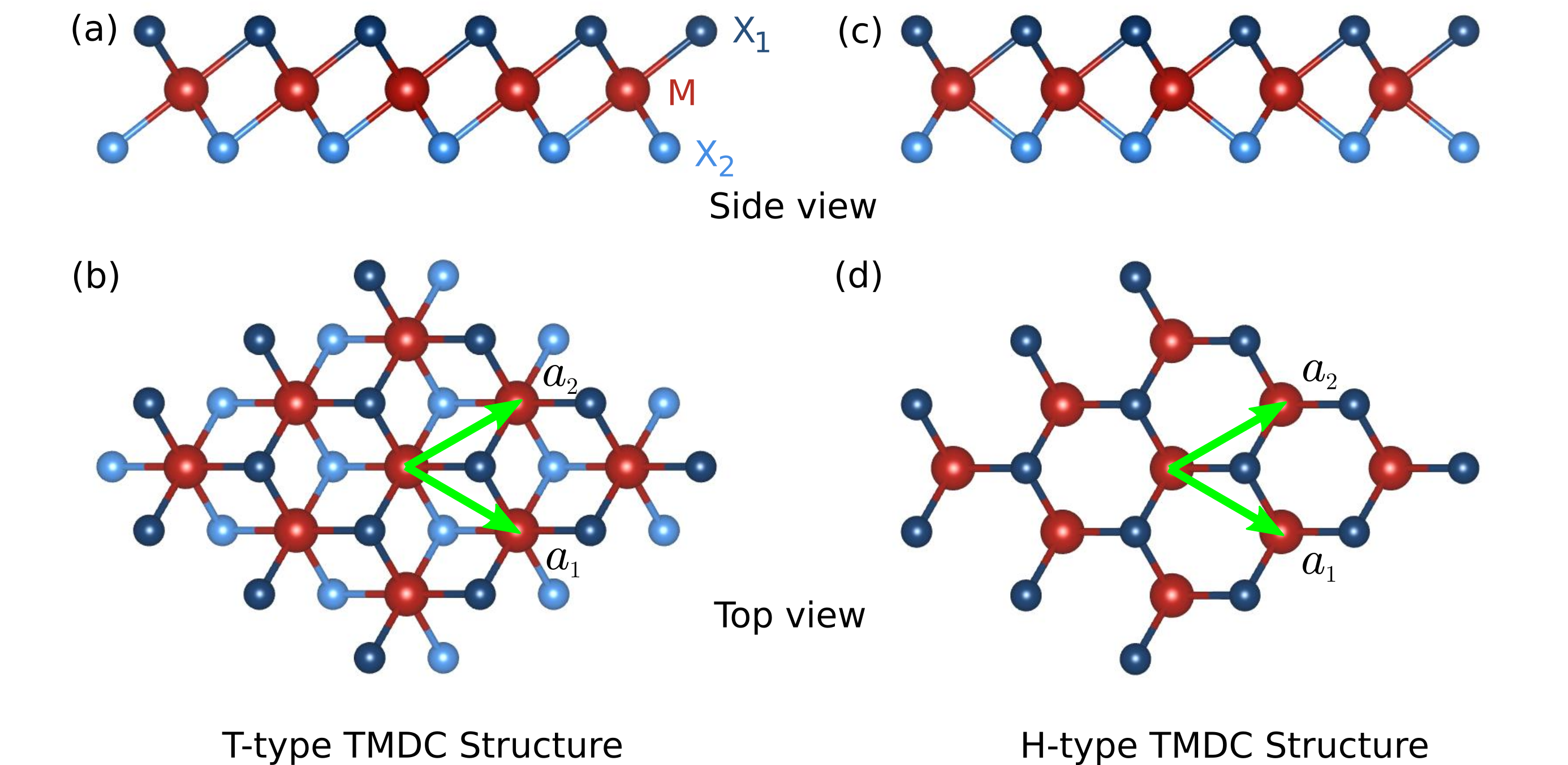}   
\caption{T-type TMDC crystal structure seen from the side (a) and top (b). For comparison we show the H-type TMDC crystal in (c) and (d) from the side
and top, respectively. $\vec{a}_1$ and $\vec{a}_2$
are the primitive lattice vectors of the 2D crystal (green arrows).}
\label{fig:Tstruct}
\end{figure}

Transition metal dichalcogenides (TMDCs) are an interesting class of materials formed by a layer of transition metal atoms (M) sandwiched between layers of chalcogens (X) with the chemical formula MX$_2$, as shown in Figure~\ref{fig:Tstruct}. TMDCs built from group-VI transition metals, such as Mo and W, generally take on the H-structure in which the transition metal exhibits trigonal prismatic coordination by the chalcogens (point group $D_{3h}$), as shown in Figure~\ref{fig:Tstruct}(c)-(d). These materials, which are usually semiconductors, have become the subject of investigation for the past several years. TMDCs can also crystallize in the T-structure, in which the metal atom is octahedrally coordinated by the chalcogens (point group $D_{3d}$), as shown in Figure~\ref{fig:Tstruct}(a)-(b). For TMDCs containing group-IV or V transition metals, the T-structure is typically metallic and either more stable or very close in energy to the H-phase. These T-type TMDCs, or T-TMDCs, are gaining attention because of their rich quantum phases. For example, with decreasing temperature, 1T-TaS$_2$ transitions from a high temperature normal metallic phase to an incommensurate CDW, near-commensurate CDW, and finally a commensurate CDW with Star-of-David clusters of Ta atoms with a $\sqrt{13} \times \sqrt{13}$ periodicity that gives rise to a correlated Mott insulator phase~\cite{tsen2015structure}. Under pressure, the Mott insulator phase melts and superconductivity develops around 2.5 GPa with $T_c$ saturating at 5 K under high pressure~\cite{sipos2008mott}. In addition, there is a suggestion that TaS$_2$ should be considered a quantum spin liquid~\cite{TaS2_QSL}.

In these T-type TMDCs, the existence of various electronic phases provides a platform to study their mutual competition and quantum criticality. Empirical tight-binding models have been constructed for the Ti based T-TMDCs~\cite{murray1972band} and for TaS$_2$~\cite{smith1985band}.  However, theoretical models for these T-type TMDCs are scarce compared to their H-type counterparts. In this work we provide comprehensive {\it ab initio} tight-binding Hamiltonians (TBH) for 9 different T-type TMDCs, considering the $d$ orbitals from transition metal atoms and the $p$ orbitals from the chalcogens. These Hamiltonians are extracted from the Wannier transformation of density functional theory (DFT) calculations without parameter fitting. In-plane strain is also included in the Hamiltonian modeling, capturing the coupling between electrons and long wavelength acoustic phonon modes. This provides a simple ``parent" electronic band structure in which additional perturbations from CDW order or other deformations can be introduced, as long as the atomic displacements from the parent structure are not too large. The application of symmetry group analysis simplifies the modeling and elucidates the nature of the various coupling terms. 

This paper is organized as follows: In Sec.~\ref{sec:props} we survey some of the notable properties of the T-TMDC materials we have modeled. In Sec.~\ref{sec:TBH} we provide details of the modeling procedure and give the explicit symmetry-constrained form of the tight-binding Hamiltonian that includes strain perturbations. Validation and analysis is provided in Sec.~\ref{sec:validation}, followed by applications to three model systems in Sec.~\ref{sec:applications}, and finally our conclusions in Sec.~\ref{sec:conclusion}. In the Appendix we tabulate the material-dependent numerical parameters extracted from DFT calculations for these tight-binding Hamiltonians.

\section{T-type TMDC Material Properties} \label{sec:props}

\begin{figure*}
\centering
\includegraphics[width=0.40\textwidth]{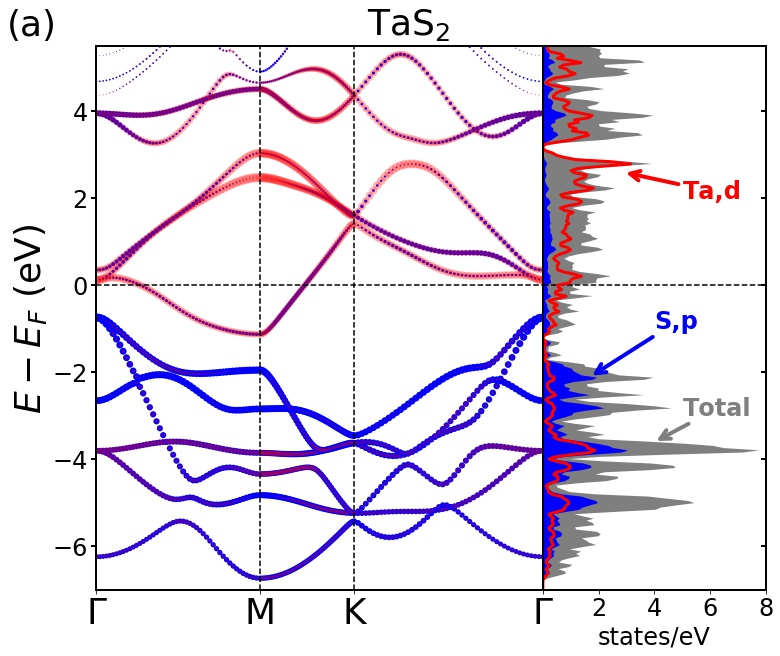}
\includegraphics[width=0.40\textwidth]{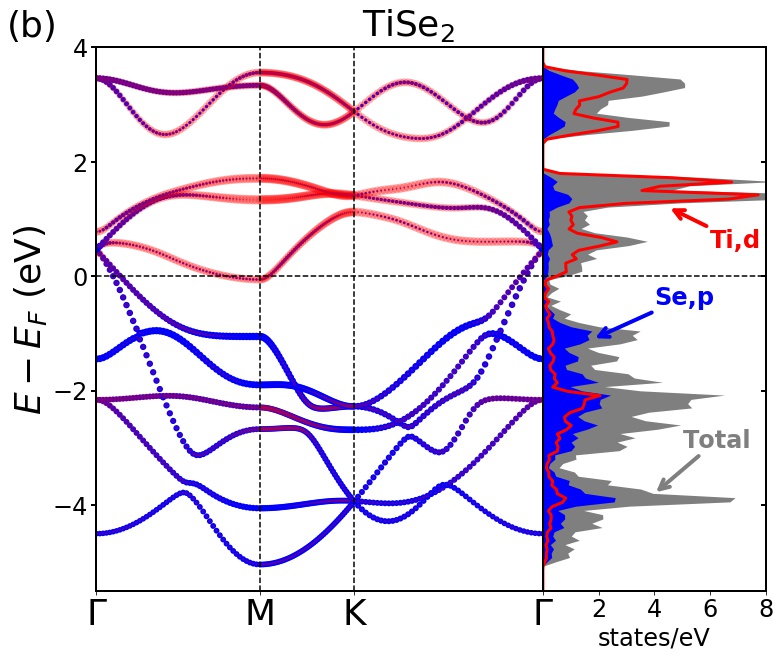}
\caption{DFT band structure and density of states (DOS) for two T-type TMDCs: (a) TaS$_2$ and (b) TiSe$_2$, showing the 11 bands nearest the Fermi level. The valence band character is dominated by the chalcogen $p$-orbitals, while the transition metal $d$-orbitals dominate the conduction band character. A notable feature is the large peak in the conduction band DOS for TiSe$_2$ which is absent for TaS$_2$.}
\label{FIG:bnd-dos}
\end{figure*}

Strain-dependent tight-binding models for group-VI TMDCs have been derived in previous work~\cite{fang2018electronic}. For these materials, specifically MX$_2$ with M = (Mo, W) and X = (S, Se), the H-structure is more stable than the T-structure, and most are semiconductors. In contrast, here we study group-IV and group-V TMDCs, with M = (Ti, Nb, Ta) and X = (S, Se, Te), where the H and T phases are very close in energy, as shown in Table~\ref{tab:Tsummary}. The T-phase of the Ti compounds has a lower ground state energy than the H-phase, and for the group-V metals (Nb and Ta) the T-phase is within 100 meV of the H-phase.
The values in Table~\ref{tab:Tsummary} are the calculated energies of the materials in the undistorted T-structure, which have metallic band structures; examples are shown in Figure~\ref{FIG:bnd-dos}. 
The energies shown in Table~\ref{tab:Tsummary} represent pristine unit cells in vacuum; in experimental contexts many factors, such as the presence of CDW distortions, substrates, finite temperature, chemical environment and intercalation, strain, and pressure, will affect the stability or metastability of a given phase~\cite{MoS2_HT_pattern,Li_MoS2,TMDC_structure,nourbakhsh2016mos2}. The T-TMDCs we study are noteworthy for the diversity of CDW patterns which can both stabilize the structure and  change the electronic properties. Table~\ref{tab:Tsummary} lists the various CDW patterns that have been observed in monolayer or bulk samples. Further, because the H-type to T-type transition can occur via a shift of a chalcogen layer, the kinetic barrier between the phases can be small enough for \emph{in situ} manipulation; for instance, an STM tip was shown to reversibly control a transition between H- and T-type structures in NbSe$_2$~\cite{Bischoff2017}.

\begin{table}[h!]
  \centering
\caption{Summary of the single-layer T-type TMDCs considered, showing the calculated energy difference between single layers of the pristine T- and H-phases, $\Delta E = E_{\mathrm{T}}-E_{\mathrm{H}}$, in units of eV per formula unit (MX$_2$), and the lattice distortions that have been observed in the various T-phase crystals as multiples of the primitive unit cell, with references where these phases were observed. The last column contains the work function, $\Delta\Phi = E_\mathrm{vac}-E_\mathrm{F}$ (in eV), discussed later in Sec.~\ref{sec:TBH}.D, for the relaxed, unstrained lattices.}
 \begin{tabular}{|l|c|c|c|}
  \hline
    Material & $\Delta E$ (eV) & T-structure CDW & $\Delta\Phi$ (eV) \\
    \hline\hline
    TiS$_2$ & $-0.43$ & $-$ & 5.64 \\
    TiSe$_2$ & $-0.36$ & 2$\times$2~\cite{di1976electronic} & 5.22 \\
    TiTe$_2$ & $-0.31$ & 2$\times$2~\cite{chen2017emergence} & 4.70 \\
    \hline
    NbS$_2$ & 0.10 & $-$ & 5.20 \\
    NbSe$_2$ & 0.09 & $\sqrt{13}\times\sqrt{13}$~\cite{nakata2016monolayer} & 4.80 \\
    NbTe$_2$ & 0.00 & 3$\times$1~\cite{battaglia2005fermi} & 4.51 \\
    \hline
    TaS$_2$ &  0.07 & $\sqrt{13}\times\sqrt{13}$~\cite{*[{}] [{ reprinted as }] wilson1975charge,*wilson2001charge} & 4.95 \\
    TaSe$_2$ & 0.07 & $\sqrt{13}\times\sqrt{13}$~\cite{wilson1974charge} & 4.57 \\
    TaTe$_2$ & 0.00 & 3$\times$1, 3$\times$3~\cite{feng2016charge} & 4.32 \\
    \hline
  \end{tabular}
  \label{tab:Tsummary}
\end{table} 

Because of these diverse CDW orders and the interplay with superconductivity, group-IV and V TMDCs have been the subject of much recent research. In the following we provide a brief summary of the most interesting features of each of type of material as motivation for the strain-dependent electronic structure modeling described in the next section. More comprehensive reviews and additional references can be found in~\cite{mcdonnell2016atomically,manzeli20172d,Voiry2015}.

TiSe$_2$ is intriguing because it exhibits a $2\times2$ CDW, a possible excitonic insulating phase~\cite{TiSe2_cercellier2007evidence}, and, with doping to suppress the CDW state, superconductivity~\cite{TiSe2_morosan2006superconductivity,TiSe2_li2016controlling}.
As a result, the CDW phase has been the subject of study in relation to electronic, excitonic, and vibrational structure effects, and much ongoing investigation focuses on the nature of the CDW transition~\cite{TiSe2_hellgren2017critical,TiSe2_fang1997bulk,TiSe2_chen2016hidden}.
Single-layer TiTe$_2$ also exhibits a $2\times2$ CDW, but it is not observed in samples with 2 or more layers~\cite{chen2017emergence}. Previous theoretical work has shown that single-layer TiS$_2$ is a semimetal with increasing band overlap under compressive strain, but switches to a semiconductor with a gap that grows with tensile strain~\cite{xu2015strain}.

NbS$_2$ and NbSe$_2$ have proven to be challenging to produce in the 1T-structure. There is a single report of experimental production of the 1T polytype of NbS$_2$ using atmospheric pressure chemical vapor deposition~\cite{carmalt2004formation}. Recently, Nakata \emph{et al.}~\cite{nakata2016monolayer} have successfully grown single-layer NbSe$_2$ on bilayer graphene and shown that the phase, T \emph{vs} H, can be controlled by the substrate temperature during growth. It has also been demonstrated that the interaction between an STM tip and the sample can induce a reversible 2H to 1T phase transition in NbSe$_2$~\cite{Bischoff2017}. On the other hand, bulk NbTe$_2$ takes on the 1T-structure but with a significant $3\times1$ CDW distortion~\cite{battaglia2005fermi}, and also exhibits superconductivity at temperatures below 1 K~\cite{van1967some}. The niobium-based TMDCs are $d^1$ materials, so the simplest ionic picture would predict metallic behavior with a single half-filled band crossing the Fermi surface. However, ARPES measurements on NbSe$_2$~\cite{nakata2016monolayer} and NbTe$_2$~\cite{battaglia2005fermi} do not show any quasiparticle crossings.

The tantalum-based TMDCs are perhaps best known for their rich CDW phases competing with superconductivity. TaS$_2$ and TaSe$_2$ are formally $d^1$ materials and are indeed metallic at high temperatures, but as the temperature is lowered they develop incommensurate, nearly-commensurate, and finally a commensurate $\sqrt{13}\times\sqrt{13}$ CDW phase. This latter phase is a correlated Mott insulator~\cite{tsen2015structure}. When the CDW phase is suppressed by either chemical doping, applying high pressure, or chalcogen substitution, superconductivity can arise at temperatures of a few K~\cite{sipos2008mott,li2012fe,liu2013superconductivity,liu2013superconductivity,liu2016nature}. On the other hand, TaTe$_2$ exhibits similar behavior to NbTe$_2$ at room temperature, including a $3\times1$ CDW reconstruction, but at temperatures below $\sim$170 K it displays a competing $3\times3$ CDW~\cite{feng2016charge} and no superconductivity~\cite{battaglia2005fermi}.

\section{\label{sec:TBH}TIGHT-BINDING HAMILTONIAN FOR THE MONOLAYER WITH STRAIN}

\subsection{Numerical Methods}

DFT calculations were performed using the VASP code~\cite{kresse1996efficient,kresse1996efficiency} with the PBE
exchange-correlation functional~\cite{perdew1996generalized}. Projector augmented wave (PAW) pseudopotentials included 6 valence electrons ($s^2p^4$) for the chalcogens and 11, 12, and 13 valence electrons for Ta ($5p^{6}6s^{1}5d^{4}$), Ti ($3s^{2}3p^{6}4s^{1}3d^{3}$), and Nb ($4s^{2}4p^{6}5s^{1}4d^{4}$), respectively. A plane-wave
energy cutoff of 420 eV was used for all materials, along with van der Waals corrections using the zero-damping DFT-D3 method~\cite{grimme2010consistent}. We employed a $\Gamma$-centered 25$\times$25$\times$1 k-point grid. All structural relaxations were carried out until the Hellman-Feynman forces on each atom had a magnitude below 0.001 eV/\AA{}. The maximally localized Wannier transformations were performed using the Wannier90 code~\cite{mostofi2008wannier90}.

\subsection{General Formulation of Strained Crystal Lattices}

As viewed from above in Fig. \ref{fig:Tstruct} (b), the TMDCs T-structure consists of 3 interpenetrating triangular lattices. Seen from the side, the transition metal lattice is sandwiched between chalcogen lattices above and below, in such a way that the metal atom is octahedrally coordinated. To model a monolayer T-TMDC we choose the primitive lattice vectors to be $\vec{a}_1=\frac{\sqrt{3}}{2}a\hat{x}-\frac{1}{2}a\hat{y}$ and $\vec{a}_2=\frac{\sqrt{3}}{2}a\hat{x}+\frac{1}{2}a\hat{y}$. The metal atom is located at the origin and the horizontal position of the upper (lower) chalcogen atom is taken to be $\pm(\vec{a}_1+\vec{a}_2)/3$.


Effects of in-plane strain on the electronic structure can be studied in the same way as they were for H-type TMDCs~\cite{fang2018electronic}, namely by displacing the crystal coordinates according to a vector field $\vec{u} = (u_x(x,y), u_y(x,y))$, where $\vec{r} = (x,y)$ gives the coordinates in the unstrained crystal and the location in the strained crystal is given by $\vec{r} + \vec{u}$. Since no physical change arises from a constant displacement, the strain field is characterized by the symmetrized derivative of $\vec{u}$ and written:
\[ u_{ij} = \frac{1}{2} \left(\partial_i u_j + \partial_j u_i\right) \]
with $i,j=x,y$. Under the $C_{3v}$ symmetry group of the crystal the strain field can be decomposed into a scalar part, $u_{xx}+u_{yy}$, representing biaxial isotropic strain, and a doublet, $(u_{xx}-u_{yy}, -2u_{xy})$, representing anisotropic strain and shear. The antisymmetrized derivative $\omega_{xy} = \frac{1}{2}\left(\partial_x u_y - \partial_y u_x\right)$ is also a scalar under rotations and represents a local rotation of the coordinates by an angle $\omega_{xy}$.

By calculating the electronic structure of the crystal subjected to different amounts of uniform strain ($u_{ij}=$ constant), we can then approximate the response to a slowly varying strain field by assuming a constant strain over any small region, referred to as a ``local strain approximation". In this manner we can study the interaction between the electronic structure and long-wavelength acoustic phonons.

The first step, considering only the physical structure, is to use DFT to optimize the lattice constant for the unstrained crystal. From the optimized crystal we can determine the height, $h$, of the chalcogen atoms above (and below) the plane of the metal atoms by relaxing the internal coordinates of a cell where the components of the lattice vectors, $\vec{a}_i$ for $i=1,2,3$, have been modified by the constant strain field as follows:
\begin{eqnarray}
a'_{ix} &=& a_{ix} + u_{xx} a_{ix} + u_{yx} a_{iy} \\
a'_{iy} &=& a_{iy} + u_{xy} a_{ix} + u_{yy} a_{iy} \\
a'_{iz} &=& a_{iz}.
\end{eqnarray}
As a function of the isotropic strain, the height $h$ of the chalcogens above and below the metal layer can be parametrized as
\begin{equation}
\label{eqn:h}
    h = d_0 - d_1 (u_{xx} + u_{yy}).
\end{equation} 
Our results for the lattice constants, $a$, unstrained chalcogen distances, $d_0$, and coefficient $d_1$, are given in Table~\ref{tab:a}. Several general trends are readily apparent in the results, in particular the increase in both lattice constant and chalcogen height with the atomic number of the chalcogen.

\begin{table}
  \centering
  \caption{Calculated equilibrium lattice constant, $a$ (\AA), distance of the chalcogen from the basal plane in the unstrained crystal, $d_0$ (\AA{)}, and coefficient $d_1$ (\AA{}) from Eq.~(\ref{eqn:h}), representing the response of the chalcogen height to isotropic strain for the TMDC MX$_2$.}
  \begin{tabular}{|cr|c|c|c|}
  \hline
    \multicolumn{2}{|c|}{\diagbox{\ M}{\raisebox{-0.8\height}{X}}} & S & Se & Te \\ \hline
     & $a$ & 3.41 & 3.53 & 3.74 \\ \cline{3-5}
    Ti & $d_0$ & 1.42 & 1.56 & 1.75 \\ \cline{3-5}
     & $d_1$ & 0.71 & 0.76 & 0.83 \\ \hline
     & $a$ & 3.35 & 3.46 & 3.62 \\ \cline{3-5}
    Nb & $d_0$ & 1.55 & 1.68 & 1.87 \\ \cline{3-5}
     & $d_1$ & 0.70 & 0.78 & 0.80 \\ \hline
     & $a$ & 3.36 & 3.49 & 3.64 \\ \cline{3-5}
    Ta & $d_0$ & 1.54 & 1.66 & 1.85 \\ \cline{3-5}
     & $d_1$ & 0.70 & 0.75 & 0.86 \\ \hline
  \end{tabular}
  \label{tab:a}
\end{table} 

\subsection{Tight-Binding Hamiltonian with Strain}

Having obtained the structural parameters for each material under study, we use DFT calculations of the electronic structure followed by a transformation to an atomic-like basis of maximally localized Wannier functions (MLWF)~\cite{marzari1997maximally,souza2001maximally} in order to determine the tight-binding parameters for a range of input strain values. By fitting the variations in the parameters as a function of $u_{xx}$, $u_{yy}$, and $u_{xy}$ we arrive at a complete tight-binding Hamiltonian for any choice of uniform strain.

A monolayer T-type TMDC has $D_{3d}$ symmetry, which includes $xz$ mirror symmetry, C$_2$ rotation symmetry about the $y$-axis, inversion symmetry, and $\mathcal{R}_3$ rotation symmetry about the $z$-axis. The 3-fold rotational symmetry of the crystal means that hopping terms to equivalent neighbor atoms will appear differently in a Hamiltonian matrix with a fixed rectangular coordinate system. However, careful choice of reference hopping vectors and analysis of the symmetry-allowed matrix elements results in significant constraints on the form of the Hamiltonian matrix.

Our model of the electronic band structure includes 11 orbitals: five $d$-orbitals, $d_{xy}$, $d_{yz}$, $d_{x^2-y^2}$, $d_{xz}$, and $d_{z^2}$ from the metal atom at the M site (origin),  and three $p$-orbitals, $p_x$, $p_y$, and $p_z$ from each chalcogen atom, located at X$_1$ (upper) and X$_2$ (lower) sites. In what follows we use this ordering of basis orbitals to define the tight-binding Hamiltonian, including hopping terms up to 3rd nearest neighbor (3NN) and all dependence on isotropic and anisotropic strain, $(u_{xx}+u_{yy})$ and $(u_{xx}-u_{yy}, 2u_{xy})$, respectively.

The on-site energy represents the interactions between orbitals located at the same atom. At $X_1$ sites the chalcogen $p$-orbital on-site energy has the form:
\begin{widetext}
\begin{equation}
\label{eqn:TBH_onsite_X}
H=\begin{bmatrix}
    \epsilon_0 & 0 & 0 \\
    0 & \epsilon_0 & 0 \\
    0 & 0 & \epsilon_1 \\
\end{bmatrix}+(u_{xx}+u_{yy})\begin{bmatrix}
	\alpha^{(0)}_0 & 0 & 0 \\
    0 & \alpha^{(0)}_0 & 0 \\
    0 & 0 & \alpha^{(0)}_1 \\
\end{bmatrix} + (u_{xx}-u_{yy})\begin{bmatrix}
	\beta^{(0)}_0 & 0 & \beta^{(0)}_1 \\
    0 & -\beta^{(0)}_0 & 0 \\
    \beta^{(0)}_1 & 0 &  0\\
\end{bmatrix}+(2u_{xy})\begin{bmatrix}
	0 & \beta^{(0)}_0 & 0 \\
    \beta^{(0)}_0 & 0 & -\beta^{(0)}_1 \\
    0 & -\beta^{(0)}_1 & 0 \\
\end{bmatrix}
\end{equation}
At X$_2$ sites, the on-site Hamiltonian is the same as the ones at X$_1$ sites (including strain). At M sites the metal $d$-orbital on-site Hamiltonian reads:
\begin{equation}
\label{eqn:TBH_onsite_M}
\begin{split}
H=&\begin{bmatrix}
    \epsilon_2 & \epsilon_5 & 0 & 0 & 0 \\
    \epsilon_5 & \epsilon_3 & 0 & 0 & 0 \\
    0 & 0 & \epsilon_2 & -\epsilon_5 & 0 \\
    0 & 0 & -\epsilon_5 & \epsilon_3 & 0 \\
    0 & 0 & 0 & 0 & \epsilon_4 \\
\end{bmatrix}+(u_{xx}+u_{yy})\begin{bmatrix}
    \alpha^{(0)}_2 & \alpha^{(0)}_5 & 0 & 0 & 0 \\
    \alpha^{(0)}_5 & \alpha^{(0)}_3 & 0 & 0 & 0 \\
    0 & 0 & \alpha^{(0)}_2 & -\alpha^{(0)}_5 & 0 \\
    0 & 0 & -\alpha^{(0)}_5 & \alpha^{(0)}_3 & 0 \\
    0 & 0 & 0 & 0 & \alpha^{(0)}_4 \\
\end{bmatrix} \\
 &+(u_{xx}-u_{yy})\begin{bmatrix}
    \beta^{(0)}_2 & \beta^{(0)}_4 & 0 & 0 & 0 \\
    \beta^{(0)}_4 & \beta^{(0)}_3 & 0 & 0 & 0 \\
    0 & 0 & -\beta^{(0)}_2 & \beta^{(0)}_4 & \beta^{(0)}_5  \\
    0 & 0 & \beta^{(0)}_4 & -\beta^{(0)}_3 & \beta^{(0)}_6 \\
    0 & 0 & \beta^{(0)}_5 & \beta^{(0)}_6 & 0  \\
\end{bmatrix}+(2u_{xy})\begin{bmatrix}
    0 & 0 & \beta^{(0)}_2 & -\beta^{(0)}_4 & \beta^{(0)}_5 \\
    0 & 0 & \beta^{(0)}_4 & -\beta^{(0)}_3 & -\beta^{(0)}_6 \\
    \beta^{(0)}_2 & \beta^{(0)}_4 & 0 & 0 & 0 \\
    -\beta^{(0)}_4 & -\beta^{(0)}_3 & 0 & 0 & 0 \\
    \beta^{(0)}_5 & -\beta^{(0)}_6 & 0 & 0 & 0 \\
\end{bmatrix}
\end{split}
\end{equation}

Each atom has 6 nearest neighbors (1NN), 3 from each of the other two sites. There are three types of first-neighbor interactions: (X$_1$-M), (X$_2$-M) and (X$_2$-X$_1$). For the (X$_1$-M) interaction we take as a reference bond the hopping from M at the origin to X$_1$ at $(\vec{a}_1+\vec{a}_2)/3$. The corresponding Hamiltonian is:
\begin{equation}
\label{eqn:TBH_1N_XM}
\begin{split}
H=&\begin{bmatrix}
    0 & 0 & t^{(1)}_0 & t^{(1)}_1 & t^{(1)}_2 \\
    t^{(1)}_3 & t^{(1)}_4 & 0 & 0 & 0 \\
    0 & 0 & t^{(1)}_5 & t^{(1)}_6 & t^{(1)}_7 \\
\end{bmatrix}+(u_{xx}+u_{yy}) \begin{bmatrix}
    0 & 0 & \alpha^{(1)}_0 & \alpha^{(1)}_1 & \alpha^{(1)}_2 \\
    \alpha^{(1)}_3 & \alpha^{(1)}_4 & 0 & 0 & 0 \\
    0 & 0 & \alpha^{(1)}_5 & \alpha^{(1)}_6 & \alpha^{(1)}_7 \\
\end{bmatrix} \\
  & +(u_{xx}-u_{yy}) \begin{bmatrix}
    0 & 0 & \beta^{(1)}_0 & \beta^{(1)}_1 & \beta^{(1)}_2 \\
    \beta^{(1)}_3 & \beta^{(1)}_4 & 0 & 0 & 0 \\
    0 & 0 & \beta^{(1)}_5 & \beta^{(1)}_6 & \beta^{(1)}_7 \\
\end{bmatrix}+(2u_{xy}) \begin{bmatrix}
    \beta^{(1)}_8 & \beta^{(1)}_9 & 0 & 0 & 0 \\
    0 & 0 & \beta^{(1)}_{10} & \beta^{(1)}_{11} & \beta^{(1)}_{12} \\
    \beta^{(1)}_{13} & \beta^{(1)}_{14} & 0 & 0 & 0 \\
\end{bmatrix}
\end{split}
\end{equation}
For the (X$_2$-M) interaction, with the hopping from M at the origin to X$_2$ at $-(\vec{a}_1+\vec{a}_2)/3$ taken as the reference, the Hamiltonian has the same form as (X$_1$-M) with an overall ($-1$) factor. 
The final first neighbor coupling, (X$_2$-X$_1$), with reference bond chosen along the positive $x$-axis, has the form:
\begin{equation}
\label{eqn:TBH_1N_XX}
\begin{split}
H=&\begin{bmatrix}
    t^{(1)}_8 & 0 & t^{(1)}_{11} \\
    0 & t^{(1)}_{9} & 0 \\
    t^{(1)}_{11} & 0 & t^{(1)}_{10} \\
\end{bmatrix}+(u_{xx}+u_{yy})\begin{bmatrix}
	\alpha^{(1)}_8 & 0 & \alpha^{(1)}_{11} \\
    0 & \alpha^{(1)}_{9} & 0 \\
    \alpha^{(1)}_{11} & 0 & \alpha^{(1)}_{10} \\
\end{bmatrix} \\
&+(u_{xx}-u_{yy})\begin{bmatrix}
	\beta^{(1)}_{15} & 0 & \beta^{(1)}_{18} \\
    0 & \beta^{(1)}_{16} & 0 \\
    \beta^{(1)}_{18} & 0 &  \beta^{(1)}_{17} \\
\end{bmatrix}+(2u_{xy})\begin{bmatrix}
	0 & \beta^{(1)}_{19} & 0 \\
    \beta^{(1)}_{19} & 0 & \beta^{(1)}_{20} \\
    0 & \beta^{(1)}_{20} & 0 \\
\end{bmatrix}
\end{split}
\end{equation}
Each atom has 6 second neighbors (2NN) of the same type. In each case the reference bond is along the positive $y$-axis. The (X$_1$-X$_1$) coupling has the form:
\begin{equation}
\label{eqn:TBH_2N_XX1}
\begin{split}
H&=\begin{bmatrix}
    t^{(2)}_0 & t^{(2)}_{3} & t^{(2)}_{4}\\
    -t^{(2)}_{3} & t^{(2)}_{1} & t^{(2)}_{5} \\
    t^{(2)}_{4} & -t^{(2)}_{5} & t^{(2)}_{2} \\
\end{bmatrix}+(u_{xx}+u_{yy})\begin{bmatrix}
    \alpha^{(2)}_0 & \alpha^{(2)}_{3} & \alpha^{(2)}_{4}\\
    -\alpha^{(2)}_{3} & \alpha^{(2)}_{1} & \alpha^{(2)}_{5} \\
    \alpha^{(2)}_{4} & -\alpha^{(2)}_{5} & \alpha^{(2)}_{2} \\
\end{bmatrix} \\
&+(u_{xx}-u_{yy})\begin{bmatrix}
    \beta^{(2)}_0 & \beta^{(2)}_{3} & \beta^{(2)}_{4}\\
    -\beta^{(2)}_{3} & \beta^{(2)}_{1} & \beta^{(2)}_{5} \\
    \beta^{(2)}_{4} & -\beta^{(2)}_{5} & \beta^{(2)}_{2} \\
\end{bmatrix}+(2u_{xy})\begin{bmatrix}
    0 & \beta^{(2)}_{6} & \beta^{(2)}_{7}\\
    \beta^{(2)}_{6} & 0 & \beta^{(2)}_{8} \\
    -\beta^{(2)}_{7} & \beta^{(2)}_{8} & 0 \\
\end{bmatrix}
\end{split}
\end{equation}
while the (X$_2$-X$_2$) interaction has some ($-1$) phase factors compared to (X$_1$-X$_1$), as shown:
\begin{equation}
\label{eqn:TBH_2N_XX2}
\begin{split}
H&=\begin{bmatrix}
    t^{(2)}_0 & -t^{(2)}_{3} & t^{(2)}_{4}\\
    t^{(2)}_{3} & t^{(2)}_{1} & -t^{(2)}_{5} \\
    t^{(2)}_{4} & t^{(2)}_{5} & t^{(2)}_{2} \\
\end{bmatrix}+(u_{xx}+u_{yy})\begin{bmatrix}
    \alpha^{(2)}_0 & -\alpha^{(2)}_{3} & \alpha^{(2)}_{4}\\
    \alpha^{(2)}_{3} & \alpha^{(2)}_{1} & -\alpha^{(2)}_{5} \\
    \alpha^{(2)}_{4} & \alpha^{(2)}_{5} & \alpha^{(2)}_{2} \\
\end{bmatrix} \\
&+(u_{xx}-u_{yy})\begin{bmatrix}
    \beta^{(2)}_0 & -\beta^{(2)}_{3} & \beta^{(2)}_{4}\\
    \beta^{(2)}_{3} & \beta^{(2)}_{1} & -\beta^{(2)}_{5} \\
    \beta^{(2)}_{4} & \beta^{(2)}_{5} & \beta^{(2)}_{2} \\
\end{bmatrix}+(2u_{xy})\begin{bmatrix}
    0 & \beta^{(2)}_{6} & -\beta^{(2)}_{7}\\
    \beta^{(2)}_{6} & 0 & \beta^{(2)}_{8} \\
    \beta^{(2)}_{7} & \beta^{(2)}_{8} & 0 \\
\end{bmatrix}
\end{split}
\end{equation}
The Hamiltonian for the (M-M) interaction has the form:
\begin{equation}
\label{eqn:TBH_2N_MM}
\begin{split}
H=&\begin{bmatrix}
    t^{(2)}_6 & t^{(2)}_{11} & 0 & 0 & 0 \\
    t^{(2)}_{11} & t^{(2)}_7 & 0 & 0 & 0 \\
    0 & 0 & t^{(2)}_8 & t^{(2)}_{12} & t^{(2)}_{13} \\
    0 & 0 & t^{(2)}_{12} & t^{(2)}_9 & t^{(2)}_{14} \\
    0 & 0 & t^{(2)}_{13} & t^{(2)}_{14} & t^{(2)}_{10} \\
\end{bmatrix}+(u_{xx}+u_{yy}) \begin{bmatrix}
    \alpha^{(2)}_6 & \alpha^{(2)}_{11} & 0 & 0 & 0 \\
    \alpha^{(2)}_{11} & \alpha^{(2)}_7 & 0 & 0 & 0 \\
    0 & 0 & \alpha^{(2)}_8 & \alpha^{(2)}_{12} & \alpha^{(2)}_{13} \\
    0 & 0 & \alpha^{(2)}_{12} & \alpha^{(2)}_9 & \alpha^{(2)}_{14} \\
    0 & 0 & \alpha^{(2)}_{13} & \alpha^{(2)}_{14} & \alpha^{(2)}_{10} \\
\end{bmatrix} \\
    &+(u_{xx}-u_{yy}) \begin{bmatrix}
    \beta^{(2)}_9 & \beta^{(2)}_{14} & 0 & 0 & 0 \\
    \beta^{(2)}_{14} & \beta^{(2)}_{10} & 0 & 0 & 0 \\
    0 & 0 & \beta^{(2)}_{11} & \beta^{(2)}_{15} & \beta^{(2)}_{16} \\
    0 & 0 & \beta^{(2)}_{15} & \beta^{(2)}_{12} & \beta^{(2)}_{17} \\
    0 & 0 & \beta^{(2)}_{16} & \beta^{(2)}_{17} & \beta^{(2)}_{13} \\
\end{bmatrix}+(2u_{xy}) \begin{bmatrix}
    0 & 0 & \beta^{(2)}_{18} &  \beta^{(2)}_{19} & \beta^{(2)}_{20} \\
    0 & 0 & \beta^{(2)}_{21} & \beta^{(2)}_{22} & \beta^{(2)}_{23} \\
    \beta^{(2)}_{18} & \beta^{(2)}_{21} & 0 & 0 & 0 \\
    \beta^{(2)}_{19} & \beta^{(2)}_{22} & 0 & 0 & 0 \\
    \beta^{(2)}_{20} & \beta^{(2)}_{23} & 0 & 0 & 0 \\
\end{bmatrix}
\end{split}
\end{equation}
Similar to first neighbor coupling, each atom has 3 third neighbors (3NN) of each of the other two types, but with different reference bonds.
The (X$_1$-M) interaction, with reference bond pointing from the origin to $-2(\vec{a}_1+\vec{a}_2)/3$, takes the form:
\begin{equation}
\label{eqn:TBH_3N_XM}
\begin{split}
H=&\begin{bmatrix}
    0 & 0 & t^{(3)}_0 & t^{(3)}_1 & t^{(3)}_2 \\
    t^{(3)}_3 & t^{(3)}_4 & 0 & 0 & 0 \\
    0 & 0 & t^{(3)}_5 & t^{(3)}_6 & t^{(3)}_7 \\
\end{bmatrix}+(u_{xx}+u_{yy}) \begin{bmatrix}
    0 & 0 & \alpha^{(3)}_0 & \alpha^{(3)}_1 & \alpha^{(3)}_2 \\
    \alpha^{(3)}_3 & \alpha^{(3)}_4 & 0 & 0 & 0 \\
    0 & 0 & \alpha^{(3)}_5 & \alpha^{(3)}_6 & \alpha^{(3)}_7 \\
\end{bmatrix} \\
  & +(u_{xx}-u_{yy}) \begin{bmatrix}
    0 & 0 & \beta^{(3)}_0 & \beta^{(3)}_1 & \beta^{(3)}_2 \\
    \beta^{(3)}_3 & \beta^{(3)}_4 & 0 & 0 & 0 \\
    0 & 0 & \beta^{(3)}_5 & \beta^{(3)}_6 & \beta^{(3)}_7 \\
\end{bmatrix}+(2u_{xy}) \begin{bmatrix}
    \beta^{(3)}_8 & \beta^{(3)}_9 & 0 & 0 & 0 \\
    0 & 0 & \beta^{(3)}_{10} & \beta^{(3)}_{11} & \beta^{(3)}_{12} \\
    \beta^{(3)}_{13} & \beta^{(3)}_{14} & 0 & 0 & 0 \\
\end{bmatrix}
\end{split}
\end{equation}
The (X$_2$-M) interaction, with reference bond from the origin to $2(\vec{a}_1+\vec{a}_2)/3$, has the same form as the above with an overall ($-1$) factor.
Finally, the (X$_2$-X$_1$) coupling has a reference vector $-2(\vec{a}_1+\vec{a}_2)/3$, and the Hamiltonian takes the form:
\begin{equation}
\label{eqn:TBH_3N_XX}
H=\begin{bmatrix}
    t^{(3)}_8 & 0 & t^{(3)}_{11} \\
    0 & t^{(3)}_{9} & 0 \\
    t^{(3)}_{11} & 0 & t^{(3)}_{10} \\
\end{bmatrix}+(u_{xx}+u_{yy})\begin{bmatrix}
	\alpha^{(3)}_8 & 0 & \alpha^{(3)}_{11} \\
    0 & \alpha^{(3)}_{9} & 0 \\
    \alpha^{(3)}_{11} & 0 & \alpha^{(3)}_{10} \\
\end{bmatrix}+(u_{xx}-u_{yy})\begin{bmatrix}
	\beta^{(3)}_{15} & 0 & \beta^{(3)}_{18} \\
    0 & \beta^{(3)}_{16} & 0 \\
    \beta^{(3)}_{18} & 0 &  \beta^{(3)}_{17} \\
\end{bmatrix}+(2u_{xy})\begin{bmatrix}
	0 & \beta^{(3)}_{19} & 0 \\
    \beta^{(3)}_{19} & 0 & \beta^{(3)}_{20} \\
    0 & \beta^{(3)}_{20} & 0 \\
\end{bmatrix}
\end{equation}
\end{widetext}

For each of the above couplings in Eqs.~(\ref{eqn:TBH_1N_XM})-(\ref{eqn:TBH_3N_XX}) there are 2 additional symmetrically equivalent couplings related by $2\pi/3$ rotations (clockwise and counterclockwise). In order to implement these rotations on the tight-binding Hamiltonian, we use the following unitary transformations with $\phi=2\pi/3$ to implement the counterclockwise 
rotation on $p$ and $d$ subspaces, respectively:
\begin{equation}
\label{eqn:Ux}
\mathcal{U}^X(\phi)=\begin{bmatrix}
\cos\phi & \sin\phi & 0 \\
-\sin\phi & \cos\phi & 0 \\
0 & 0 & 1 \\
\end{bmatrix}
\end{equation}

\begin{equation}
\label{eqn:Um}
\mathcal{U}^M(\phi)=\begin{bmatrix}
\cos2\phi & 0 & -\sin2\phi  & 0 & 0 \\
0 & \cos\phi & 0 & -\sin\phi  & 0 \\
\sin2\phi  & 0 & \cos2\phi & 0 & 0 \\
0 & \sin\phi  & 0 & \cos\phi & 0 \\
& 0 & 0 & 0 & 1 \\
\end{bmatrix}
\end{equation}
Note that the orbital subspaces rotate in the opposite direction from the coordinate axes.
The 2NN couplings each have 3 additional symmetrically equivalent couplings related by a mirror symmetry in the $xz$-plane.

The strain-dependent tight-binding Hamiltonian for each material is specified by 163 parameters, which is significantly fewer than the 1056 parameters in a generic 11-dimensional Hamiltonian with isotropic and anisotropic strain dependence. Furthermore, these parameters are all extracted directly from the Wannier transformation of the DFT results (averaging over symmetrically equivalent terms) and are not the result of any fitting to the band structure. The parameters for each material are tabulated in the Appendix.

\subsection{Work Function}

The work function for each material is the difference in energy between the Fermi level of the TMDC and the vacuum (a location far from the TMDC layer): $\Delta\Phi = E_\mathrm{vac}-E_\mathrm{F}$. This can be easily extracted from the DFT total potential for each value of isotropic strain, and shows a mild (few percent) dependence on the strain, which can affect the diagonal $\epsilon$ values as well as the $\alpha^{(0)}$ couplings. The values reported in the tables in the Appendix have the strain dependent work function subtracted from the on-site couplings before fitting. The work function for the unstrained materials, which decreases with increasing atomic number of both the metal and the chalcogen atoms, was included in Table~\ref{tab:Tsummary}.

\subsection{Spin-Orbit Coupling}

Atomic spin-orbit coupling (SOC) is described by the Hamiltonian $H_\mathrm{soc} = \lambda_\mathrm{soc} \vec{L} \cdot \vec{S}$. We can extract the value of $\lambda_\mathrm{soc}$ for each atomic species by considering the splitting between valence states in a single atom as calculated by DFT. Assuming the wavefunctions are eigenstates of $\vec{J}$, $\vec{L}$, and $\vec{S}$, the energy splitting is given by the difference between $j=\ell+1/2$ and $j=\ell-1/2$ states, namely $\Delta E_\mathrm{soc} = \lambda_\mathrm{soc} (\ell+1/2)$. The value of $\lambda_\mathrm{soc}$, in units of eV, is shown in Table~\ref{tab:tmdc_SOC} for the 6 species considered here. 

\begin{table}[h!]
  \centering
  \caption{The atomic spin-orbit coupling strength $\lambda_\mathrm{soc}$, in units of eV.}
  \begin{tabular}{|p{17mm}|c|c|c|c|c|c|}
  \hline
  & Ti & Nb & Ta & S & Se & Te\\
  \hline
\ $\lambda_\mathrm{soc}$ (eV) & $0.018$ & $0.071$ & $0.232$ &  $0.056$ & $0.247$ & $0.512$\\
\hline
  \end{tabular}
  \label{tab:tmdc_SOC}
\end{table} 

\section{\label{sec:validation}Model Validation}

\begin{figure*}[htbp]
\centering
\includegraphics[width=0.35\textwidth]{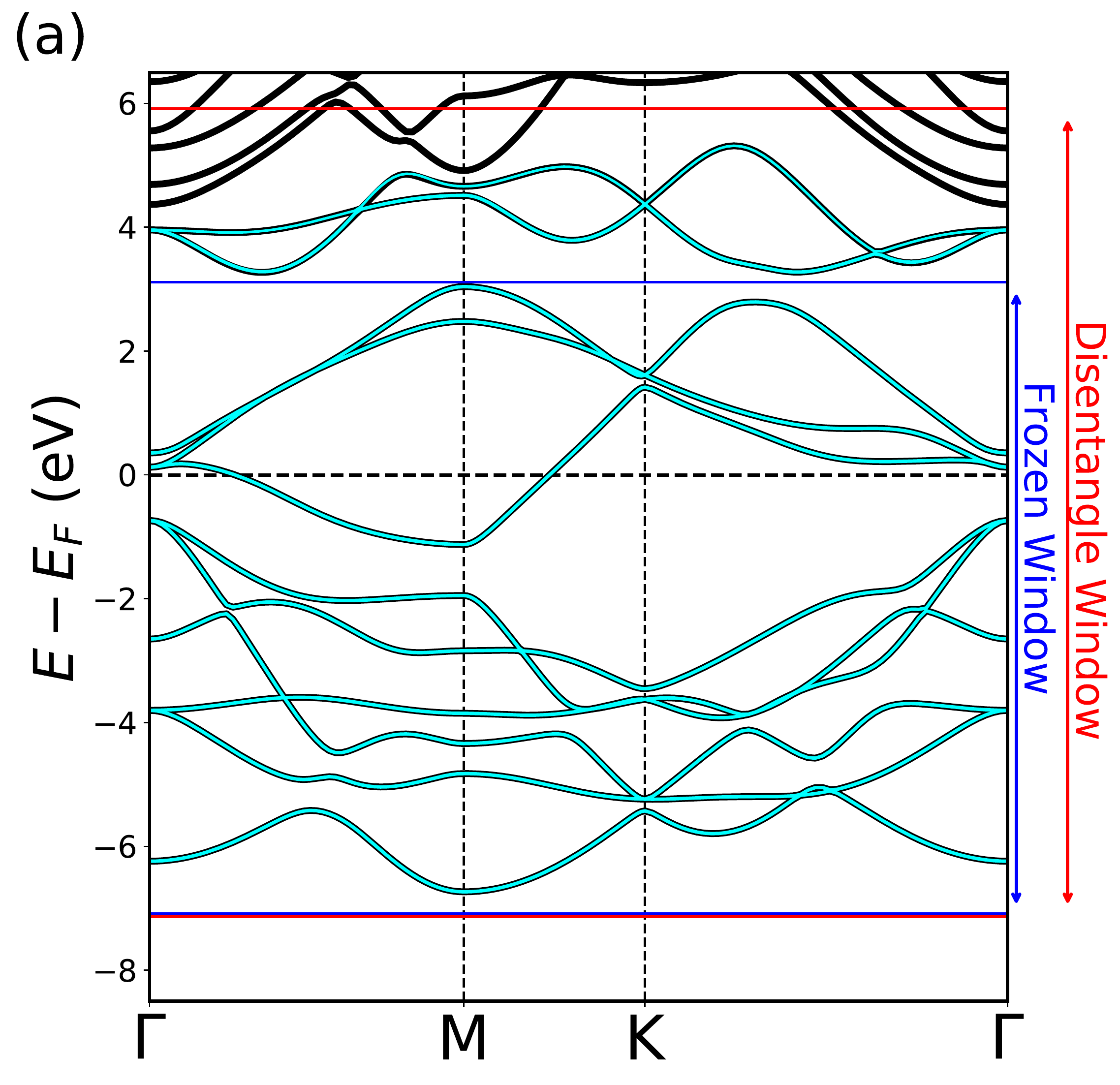}
\includegraphics[width=0.35\textwidth]{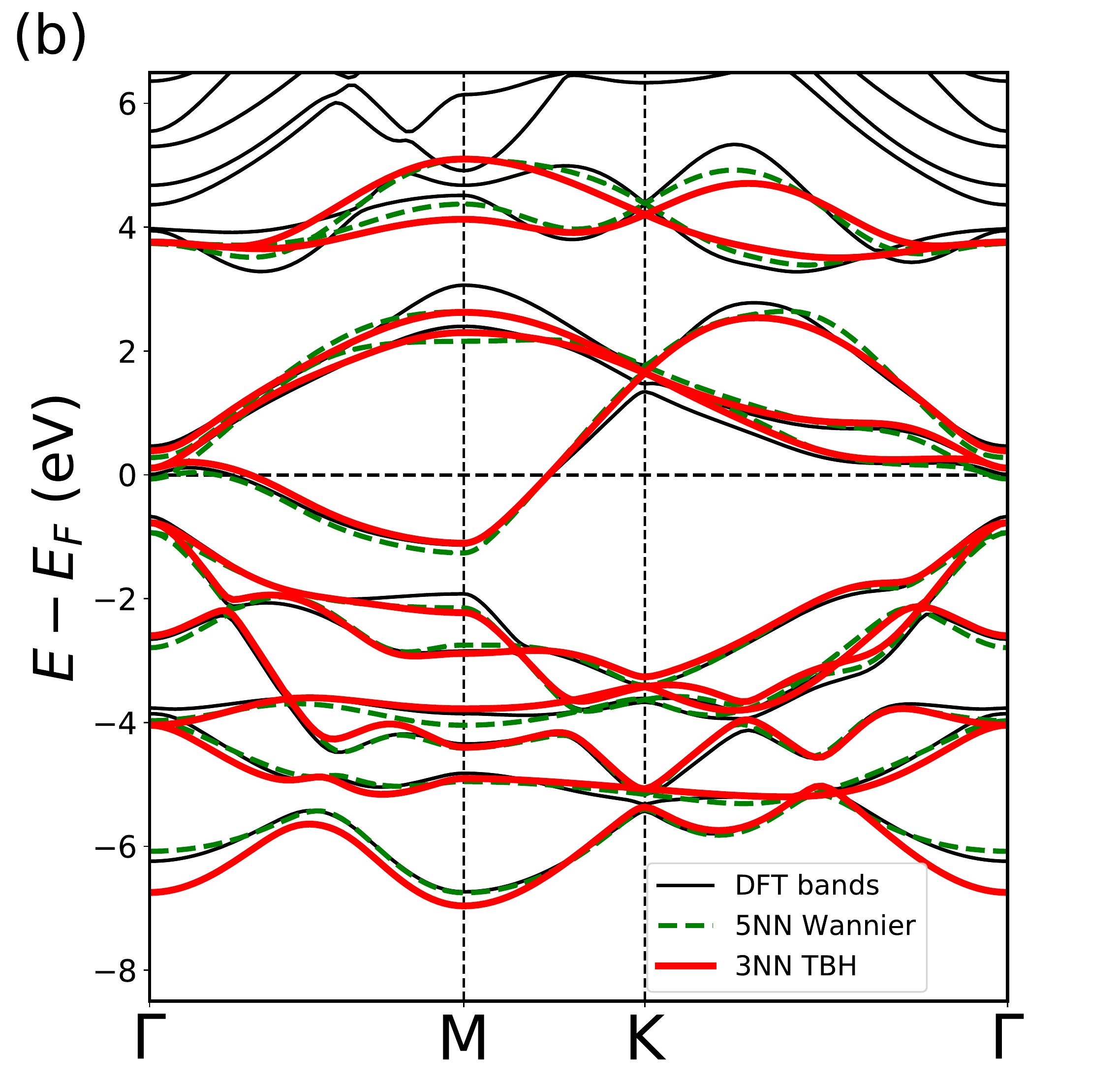}
\caption{(a) Comparison of the band structure calculated with DFT (black) and the full Wannier basis (cyan) for unstrained TaS$_2$. Also shown are the energy windows used to define the Wannier transformation and disentangle the bands arising from other orbitals~\cite{mostofi2008wannier90,souza2001maximally}. (b) Comparison of the band structure computed directly with DFT (thin black), 5NN Wannier basis (dashed green), and the 3NN TBH parametrization (thick red) for unstrained TaS$_2$.}
\label{FIG:TaS2-wannier}
\end{figure*}

In this section we demonstrate the accuracy with which our strain-dependent TBH reproduces the DFT band structures. Fig.~\ref{FIG:TaS2-wannier}(a) shows that the band structure near the Fermi energy is captured by the 11-band Wannier transformation, because the bands calculated with DFT and those reconstructed from the full basis of Wannier functions show virtually no differences.

Our tight-binding model, however, ignores interactions beyond 3NN, and it is important to examine the extent to which neglecting those longer range couplings affects the band structure. Fig.~\ref{FIG:TaS2-wannier}(b) shows the 3NN TBH band structure along with the full Wannier reconstruction and a 5NN truncation. The qualitative band structures agree well, and the most significant quantitative differences appear in bands far from the Fermi level. These differences arise from the overlap with orbitals that are not included in our 11-band model. The Wannierization procedure used to calculate the MLWF includes the effects from these other states as longer range hopping terms, beyond even 5NN.

\begin{figure*}[hbtp]
\centering
\includegraphics[width=0.3\textwidth]{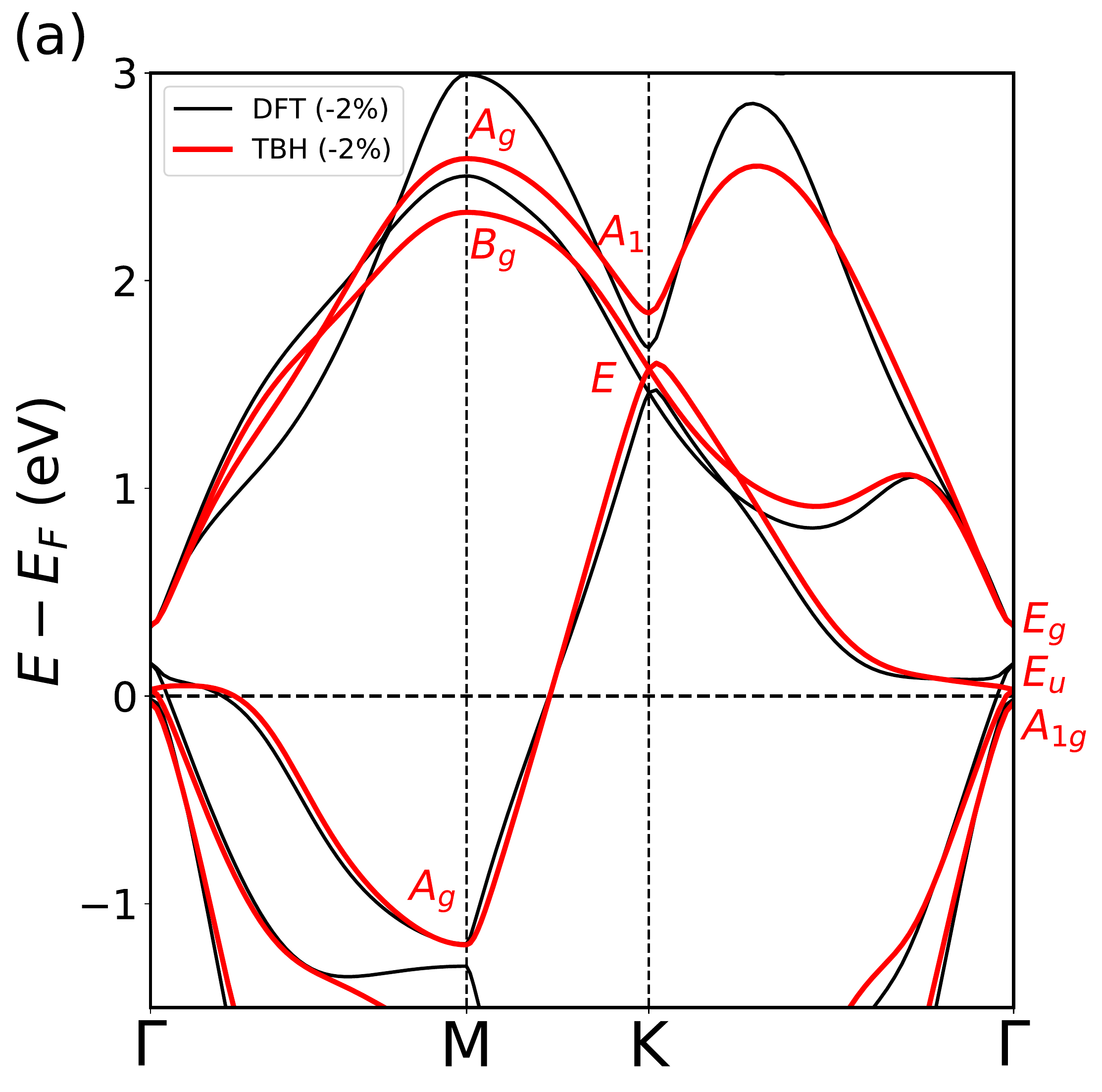}
\includegraphics[width=0.3\textwidth]{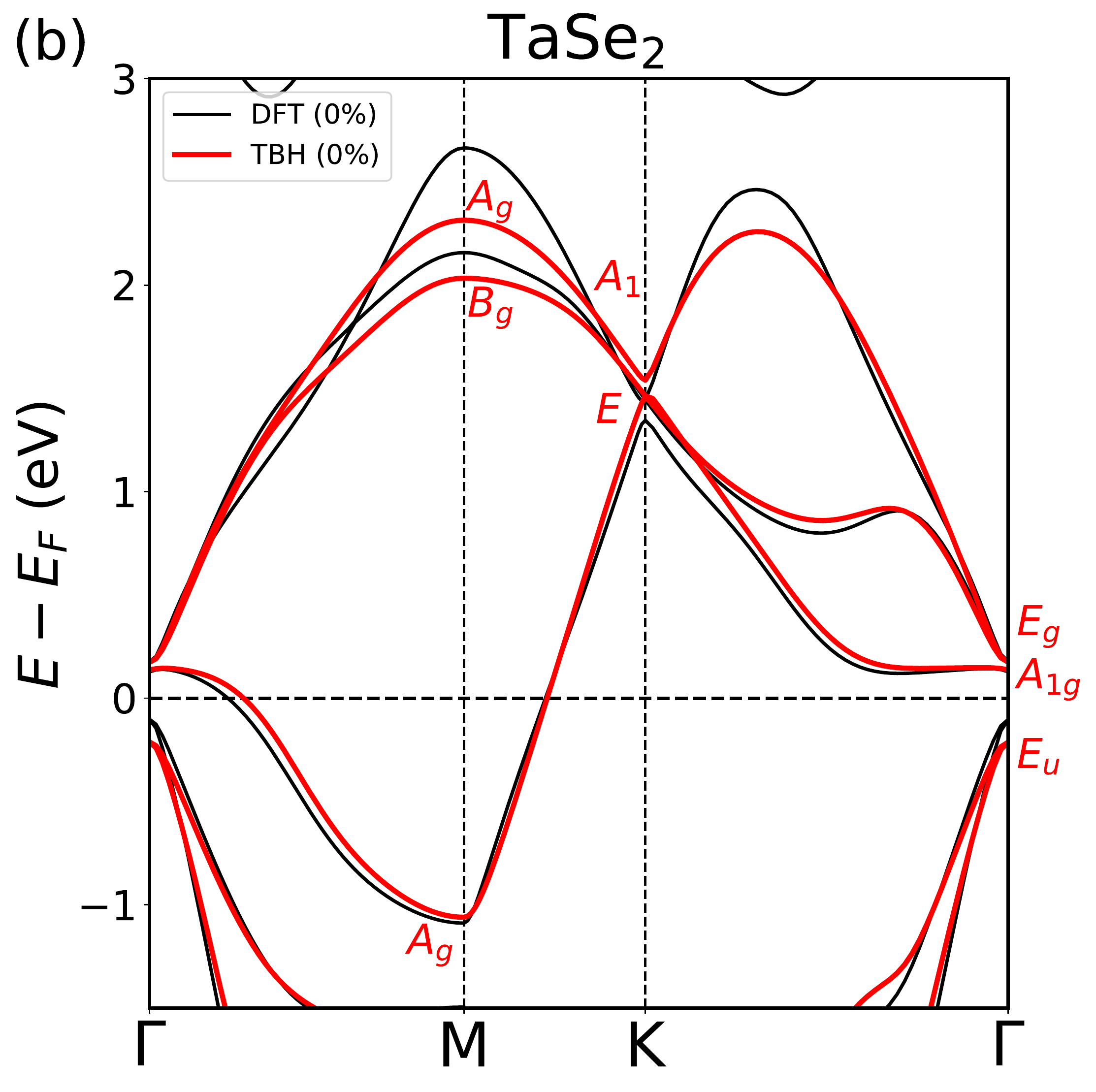}
\includegraphics[width=0.3\textwidth]{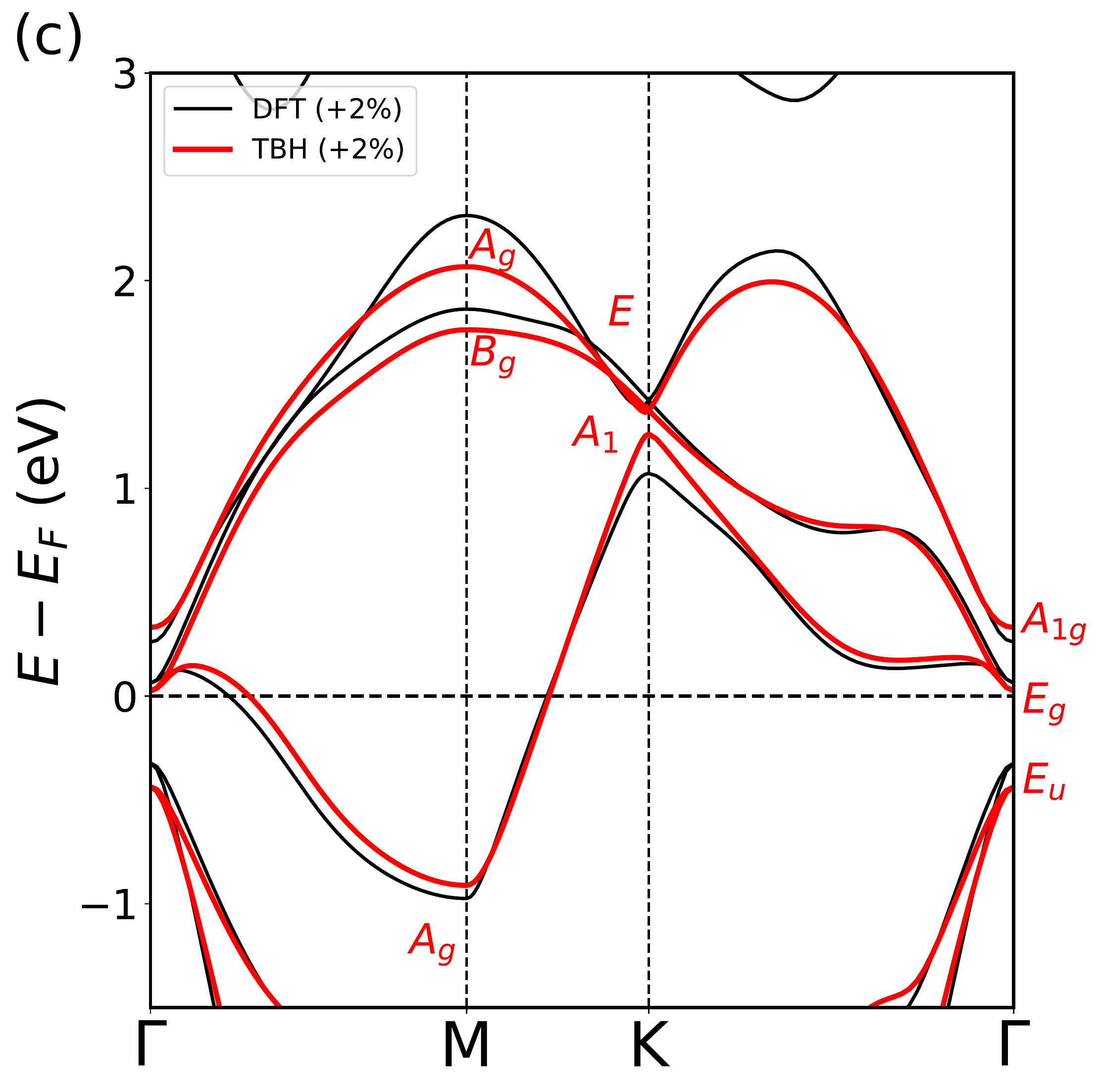}
\caption{Comparison of the band structure calculated using the 3NN TBH (red) and DFT (black) for a TaSe$_2$ crystal with three values of isotropic strain: (a) 2\% compression, (b) unstrained, and (c) 2\% expansion. The TBH bands are labeled by the irreducible representation of the symmetry group of the wave-vector at each high symmetry point. ($\Gamma$: D$_{3d}$, M: C$_{2h}$, K: D$_3$).}
\label{FIG:TaSe2-strain}
\end{figure*}

In Fig.~\ref{FIG:TaSe2-strain}, we show the effect of 2\% isotropic compression and expansion on the band structure of TaSe$_2$. The black bands are the result of DFT calculations and the red bands are from our 3NN tight-binding model, demonstrating that our model tracks the decrease in band dispersion as the crystal is expanded. From the tight-binding Hamiltonian it is easy to determine the irreducible representations of each band at the high symmetry points, $\Gamma$, M, K (see Fig.~\ref{FIG:TaSe2-strain}). Changing the lattice constant from 2\% compression to 2\% expansion results in the reordering of representations at the $\Gamma$-point near $E_F$, with the $A_{1g}$ singlet rising up through first the $E_u$ doublet (at the equilibrium lattice constant) and then the $E_g$ doublet (at 2\% expansion). This reordering is accurately captured by the tight-binding model. However, the band reordering at the K point caused by lattice expansion, with the singlet $A_1$ sinking below the $E$ doublet, has already occurred in the DFT at the equilibrium lattice constant, whereas it requires a slight expansion in the lattice before it occurs in the tight-binding model.

\begin{figure}[htbp]
\centering
\includegraphics[width=.8\columnwidth]{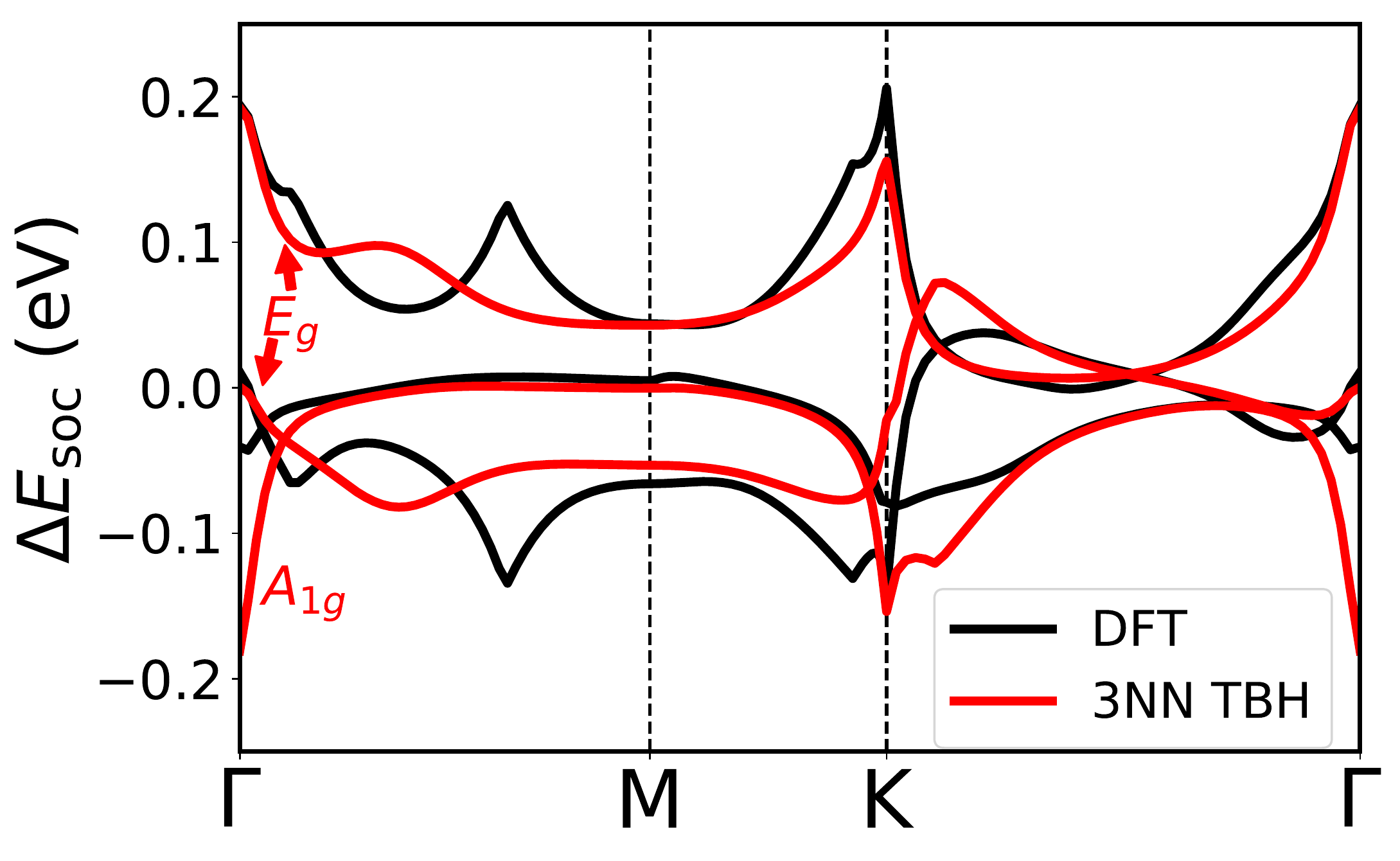}   
\caption{Change in band energy due to spin-orbit coupling for three bands near the Fermi level in unstrained TaSe$_2$. These bands can be identified as an A$_{1g}$ singlet and E$_g$ doublet at $\Gamma$. The difference in DFT calculated band energies with and without SOC, $\Delta E_\mathrm{soc} = E_\mathrm{with-soc}-E_\mathrm{no-soc}$, is shown in black, whereas the differences in tight-binding eigenvalues with and without SOC is shown in red.} 
\label{FIG:soc-compare}
\end{figure}

We can also compare the TBH with the addition of atomic spin-orbit coupling to the DFT results. Specifically, we consider the three bands in TaSe$_2$ that lie just above $E_F$ at $\Gamma$ in the unstrained crystal. Figure~\ref{FIG:soc-compare} shows the change in energy as calculated within DFT (black) or using the tight-binding parametrization including atomic SOC: there is good quantitative agreement between the two approaches.

\section{Applications}\label{sec:applications}

There are many proposals for strain engineering of 2D materials. A 2D pattern of artificial atoms has been generated by draping a single layer of MoS$_2$ over a periodic array of nanocones in a substrate, creating regions of higher and lower biaxial strain~\cite{li2015optoelectronic}. More recently, draping a graphene sheet over a step edge in a copper substrate has been shown to produce 1D ripples along the direction of the step edge~\cite{banerjee2019strain}. Lattice distortions also occur spontaneously in systems that undergo CDW reconstructions. Though these generally occur over much shorter length scales, for small distortions our strain model should be able to capture the most important effects. We demonstrate the usefulness and limitations of our strain-dependent tight-binding model for systems with non-uniform strain patterns by applying it to three example structures: a 1D sinusoidal ripple in TaS$_2$, the 2$\times$2 CDW in TiSe$_2$, and the $\sqrt{13}\times\sqrt{13}$ CDW in TaS$_2$.

\subsection{One-dimensional rippling}\label{sec:ripple}
 We have modeled the effects of long-wavelength lattice distortions on the electronic structure of TaS$_2$ subject to a sinusoidal strain of variable amplitude and wavelength in the $x$-direction. Our model accounts only for in-plane lattice distortions, which were shown to be important in reproducing the experimental data in Ref~\cite{banerjee2019strain}. Our treatment here does not consider out of plane displacements.
\begin{figure*}[htbp]
\centering
\includegraphics[width=\textwidth]{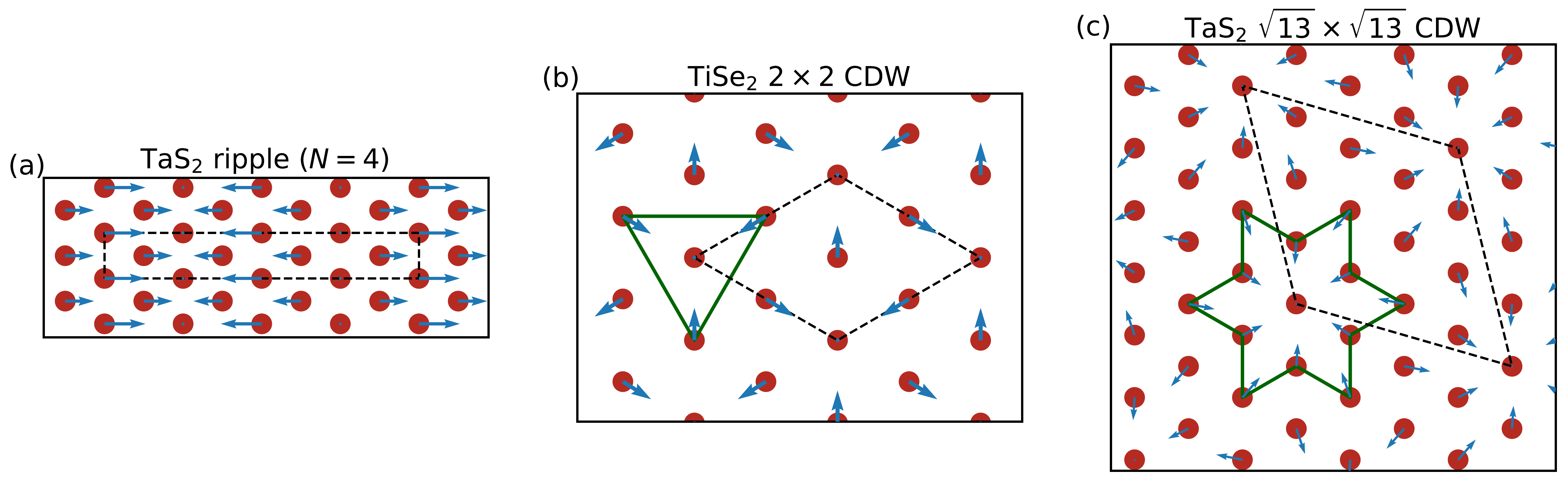}
\caption{The displacements of the metal atoms used in the applications of the strain-dependent TBH. (a) Sinusoidal strain in the $x$-direction with a period of $N=4$ rectangular units. (b) $2\times2$ CDW pattern for TiSe$_2$. (c) $\sqrt{13}\times\sqrt{13}$ CDW pattern for TaS$_2$. The supercells are shown with dashed black lines, and the chalcogen atoms are omitted for clarity.}
\label{fig:displacements}
\end{figure*}

To model a non-uniform strain pattern, we first create an atomic supercell that is large enough to include the entire periodic strain pattern. For the 1D ripple, we use a rectangular unit cell containing two formula units of TaS$_2$, and repeat it $N$ times in the $x$-direction, as shown schematically in Fig.~\ref{fig:displacements}(a) for $N=4$. Our model includes up to 3NN hopping, so each atom in the supercell interacts with 18 other atoms in addition to the on-site interaction. For each ``bond" between interacting atoms we calculate the matrix elements in the Hamiltonian using our tight-binding parametrization with the components of the strain field evaluated at the center of the bond. This approximation ensures that the model Hamiltonian remains hermitian. For the 1D ripple the displacement field is $u(r) = A\sin\left(2\pi x/a\sqrt{3}N\right) \hat{x}$ and thus $u_{xx} = B\cos\left(2\pi x/a\sqrt{3}N\right)$ and $u_{yy}=u_{xy}=u_{yx}=0$.

The supercell geometry and consequent folded band structure complicates the interpretation of the electronic structure. In this example our ripple has $2\times 11\times N$ bands. One way to make connections with the simpler pristine crystal band structure is through an ``unfolding procedure" to obtain the ``effective band structure". This method has been established to interpret the perturbations to the pristine band structure due to the presence of impurities, disordered alloys, and structural reconstructions in DFT calculations that use a supercell geometry~\cite{boykin2005practical,ku2010unfolding,popescu2012extracting,band_unfolding_disordersolid}. The effective band structure can be further compared with the band structure derived from angle-resolved photoemission spectroscopy (ARPES).

\begin{figure*}[hbtp]
\centering
\includegraphics[width=\textwidth]{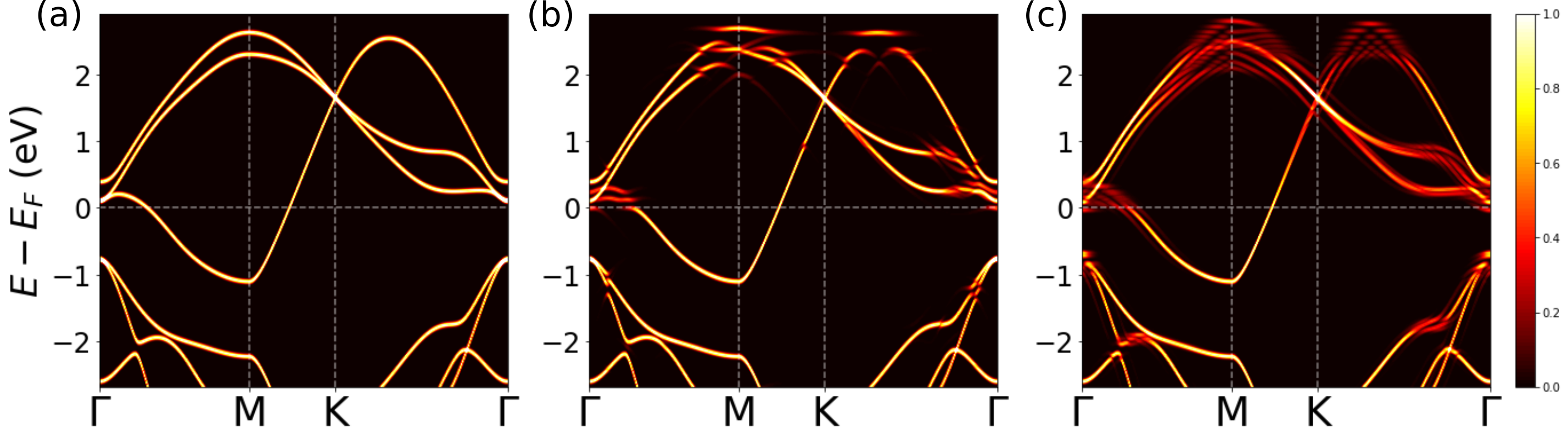}
\caption{Band structures for a $N\times$1 rectangular supercell of TaS$_2$, unfolded to a path in the Brillouin zone of the 3-atom primitive unit cell. (a) TBH bands for $N=4$ without strain. (b) TBH bands for $N=4$ subject to 1D sinusoidal strain with amplitude $B=0.02$ and wavelength $\lambda=23$ \AA{}. (c) TBH bands for $N=16$, subject to 1D sinusoidal strain with amplitude $B=0.02$ and wavelength $\lambda=93$ \AA{}.}
\label{FIG:TaS2-1d}
\end{figure*}

In practice, the unfolding procedure is carried out with the proper crystal momentum projection based on the pristine unit cell, where the corresponding Fourier component determines the unfolding weight. Figure~\ref{FIG:TaS2-1d} shows the unfolded band structures for an $N=4$ TaS$_2$ supercell both with and without the 1D sinusoidal strain. For comparison, we studied a supercell with displaced atoms directly using DFT, followed by the Wannier transformation and unfolding of the supercell band structure, and found only small differences. While in this case the DFT+Wannier calculation is not computationally prohibitive, it still takes several hours of CPU time for each configuration, and the resources required grow with the size of the supercell. By contrast, once the framework of the TBH supercell has been set up, the unfolded band structure can be produced much faster for any chosen amplitude or wavelength of the strain pattern, and for any of the 9 T-type TMDC materials we have modeled. 
Increasing the wavelength of the strain pattern to $\lambda = 93$ \AA{} ($N=16$) results in a supercell with 96 atoms, which makes the DFT calculation very computationally demanding, while the TBH result, shown in Fig.~\ref{FIG:TaS2-1d}(c), is obtained in under 1 minute using a laptop computer.


In T-TMDCs, CDW order is prevalent and affects the underlying electronic properties. Compared to the pristine crystal structure, the CDW order and electron-phonon couplings cause deformed and reconstructed atomic positions. For example, the T-TiSe$_2$ crystal exhibits a $2\times2$ reconstruction, while the T-TaS$_2$ crystal shows so-called Star-of-David deformations in the commensurate CDW phase with a $\sqrt{13} \times \sqrt{13}$ supercell. We apply our TBH model to these situations next, and compare to DFT results when possible.

\subsection{TiSe$_2$ CDW} \label{sec:TiSe2cdw}

\begin{figure*}[hbtp]
\centering
\includegraphics[width=\textwidth]{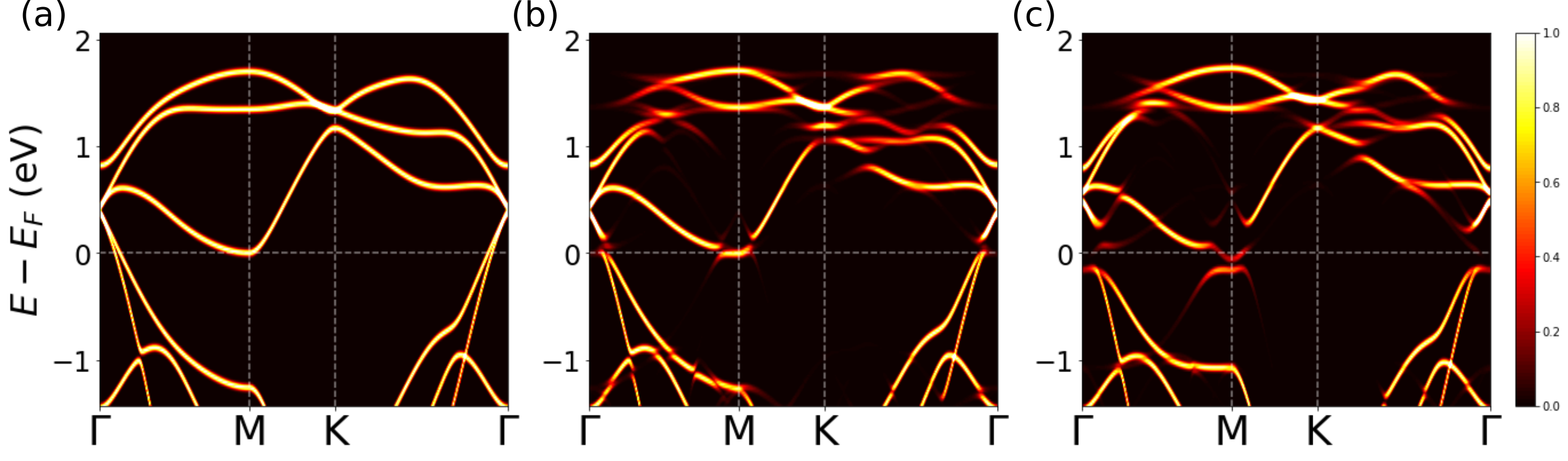}
\caption{Band structures for the 2$\times$2 CDW supercell of TiSe$_2$, unfolded to a path in the Brillouin zone of the 3-atom primitive unit cell. (a) TBH bands without the CDW distortion. (b) TBH bands for a strain pattern fit to the CDW displacements calculated with DFT. (c) DFT band structure for the relaxed CDW, for comparison.}
\label{FIG:TiSe2_unfold}
\end{figure*}

For TiSe$_2$ we first build a 2$\times$2 supercell, perturb the Ti atoms slightly, and use DFT to relax the positions of all atoms. The CDW indeed develops, with changes in the distances between the Ti atoms of $\pm 0.08$ \AA{}, which is 2.3\% of the optimized DFT lattice constant of 3.53 \AA{}. This amount of strain is at the upper limit of the uniform strain which we used to extract the parameters of the strain-dependent TBH and where the response in the TBH parameters was still quite linear. A schematic of the metal atom displacements is shown in Fig.~\ref{fig:displacements}(b).
We match a continuous displacement field to the positions of the Ti atoms in the relaxed 2$\times$2 supercell by fitting Fourier components of the first shell of reciprocal lattice vectors. The derivative of this displacement field is then used to determine the components of the strain field at the center of each bond between interacting atoms in the 2$\times$2 supercell. Unlike the case of the 1D ripple, the TiSe$_2$ CDW strain pattern includes a non-zero antisymmetric scalar, $\omega_{xy}$, which represents a local rotation. This is implemented in the TBH by rotating the strain-independent piece of the Hamiltonian for each bond by an angle $\phi=\omega_{xy}(x,y)$ evaluated at the center of each bond, using the matrices of Eqs.~(\ref{eqn:Ux}) and (\ref{eqn:Um}). Figure~\ref{FIG:TiSe2_unfold} shows the pristine and CDW-strained TiSe$_2$ bands unfolded to the primitive cell Brillouin zone. The opening of small gaps at the Fermi level is clearly captured, and the unfolded bands reproduce well those calculated directly with DFT+Wannier, shown in Fig~\ref{FIG:TiSe2_unfold}(c).

\subsection{TaS$_2$ CDW} \label{sec:TaS2cdw}

\begin{figure}[htbp]
\centering
\includegraphics[width=0.4\textwidth]{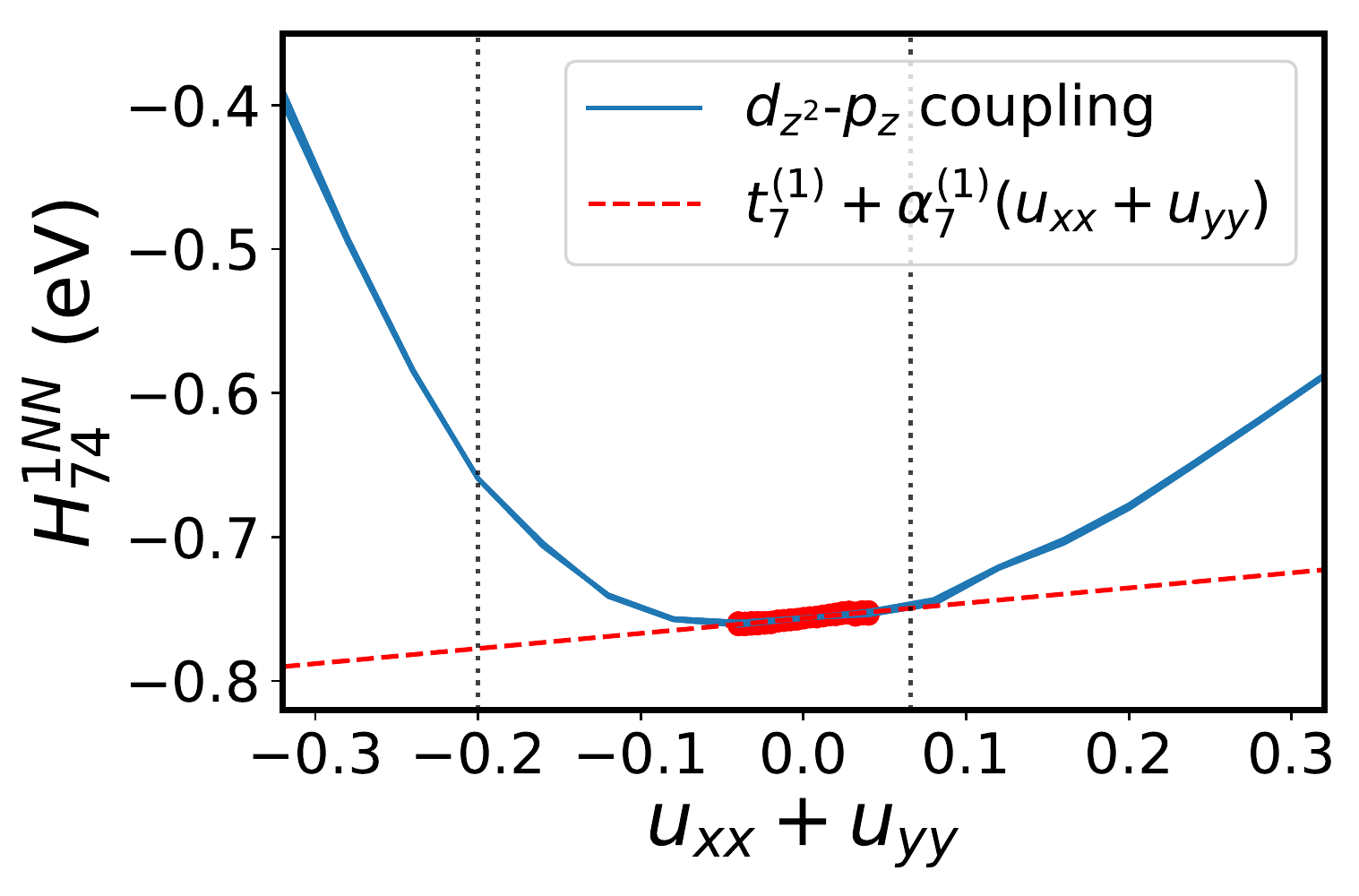}
\caption{Magnitude of the 1NN matrix element $H^{1NN}_{74}$, namely the $d_{z^2}$-$p_z$ coupling in TaS$_2$, as a function of isotropic strain, $u_{xx}+u_{yy}$. The red points represent strain in the range $\pm$2\% used to produce the linear parametrization (red dashed line) for the strain-dependent TBH. The vertical dotted black lines indicate the range of isotropic strain occurring in the TaS$_2$ CDW pattern.}
\label{FIG:nonlin}
\end{figure}

\begin{figure*}[hbtp]
\centering
\includegraphics[width=\textwidth]{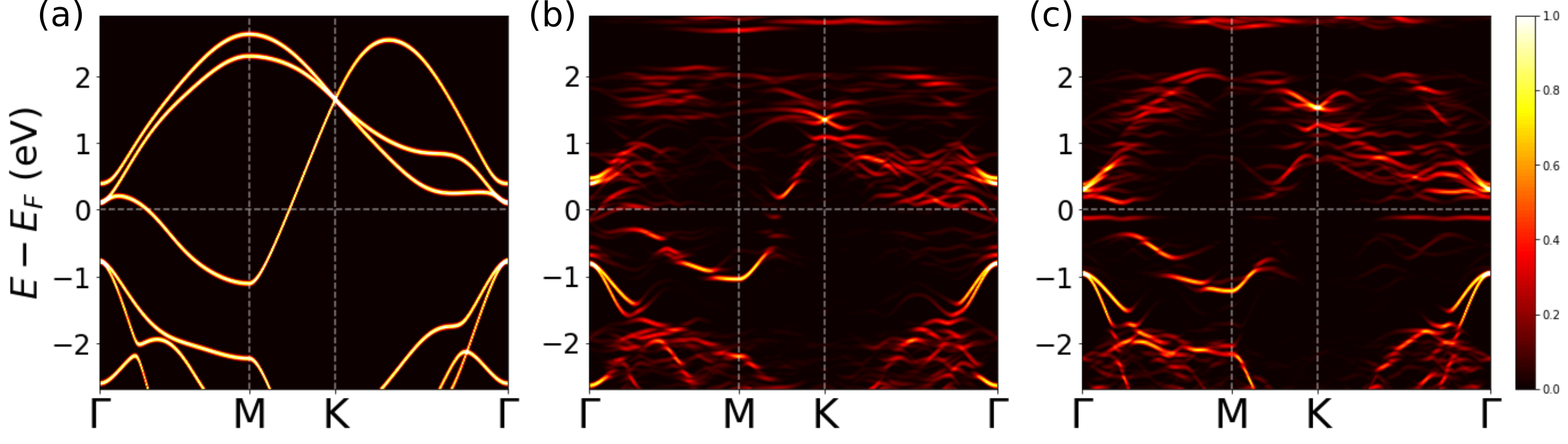}
\caption{Band structures for the $\sqrt{13}\times\sqrt{13}$ CDW supercell of TaS$_2$, unfolded to a path in the Brillouin zone of the 3-atom primitive unit cell. (a) TBH bands without the CDW distortion, (b) TBH bands for a strain pattern fit to the CDW displacements calculated with DFT, and (c) DFT band structure for the relaxed CDW.}
\label{FIG:TaS2cdw}
\end{figure*}

The simple, symmetric model we have developed reveals its limitations when applied to the $\sqrt{13}\times\sqrt{13}$ CDW pattern that occurs in TaS$_2$ and TaSe$_2$. For the former, relaxation of the supercell using DFT yields the experimentally observed CDW pattern, shown schematically in Fig.~\ref{fig:displacements}(c). The nearest Ta-Ta distances can increase by as much as 0.43 \AA{} while other distances decrease by up to 0.2 \AA{}, a range of $-6$\% to +13\% of the 3.36 \AA{} lattice constant. Displacements this large might be expected to exceed the linear strain regime parametrized in our model. Indeed, calculations using larger isotropic strains show that the response of one of the 1NN Ta-S couplings becomes very nonlinear, as shown in Fig.~\ref{FIG:nonlin}.
Using the strain-dependent TBH to compute the electronic structure for the TaS$_2$ CDW in the same manner as for TiSe$_2$, including local rotation $\omega$ and taking the strain field from a fit to the Ta positions in the relaxed DFT CDW structure, we see in Fig.~\ref{FIG:TaS2cdw} that many of the smaller features in the unfolded CDW band structure are well reproduced. However, the TBH model fails to capture the significant flat band that emerges at $\Gamma$ and is clearly visible in the DFT calculation, Fig.~\ref{FIG:TaS2cdw}(c).
In this case the perturbation from the pristine T-structure is so large that nonlinear effects in the strain response become important. The same is true for distorted phases such as the T'-structure (which has a dimerization of metal atoms in one direction). This is not a breakdown of the underlying procedure, for one can still use the Wannier transformation of the DFT results to construct accurate tight-binding models to study the coupling to acoustic phonons. However, the complexity increases significantly, preventing us from writing down simple, symmetric models.

\section{\label{sec:conclusion}CONCLUSION}
We have used the Wannier transformation of plane-wave DFT calculations to construct precise, strain-dependent tight-binding Hamiltonians for group-IV and group-V TMDCs that adopt the T-structure. We have further augmented the models with on-site spin orbit coupling terms and determined the work function for each material, providing a simple parametrization of the interaction of the electronic structure with long-wavelength acoustic phonons, which induce deformations of primitive vectors for the structural unit cell. Moreover, we have demonstrated how to model the short-wavelength CDW distortions with atomic displacements. The effects of such perturbations are visualized through the technique of band-structure unfolding. Similarly, the approach can be generalized to model crystals with internal atomic displacements due to optical phonons.

From a broader perspective, the strain response of single layers is a necessary ingredient for the construction of models of van der Waals heterostructures with multiple layers. These material systems have attracted attention recently since the discovery of unconventional correlated insulating and superconducting states in magic-angle twisted bilayer graphene~\cite{TWBLG_Mott,TWBLG_SC,TWBLG_pressure_EXP}. In terms of its electronic structure modeling, the layer deformation and strain from the mechanical relaxations in a twisted bilayer are shown to be relevant in modifying the electronic structure when compared with experimental observations~\cite{Koshino_relaxTwBLG}. Our modeling of single layers with strain paves the way for investigating heterostructures involving diverse types of T-TMDCs and the interplay between different order parameters. The other crucial ingredient necessary for a comprehensive, single-particle model of the electronic structure, is the proper interlayer couplings between orbitals in adjacent layers~\cite{MacDonald_abinitio,TwBLG_interlayer}. This will be explored in future work. 


\begin{acknowledgements}
We thank Stephen Carr, Jhih-Shih You, Dennis Huang, and Philip Kim for useful discussions. This work was supported by the STC Center for Integrated Quantum Materials, NSF Grant No. DMR-1231319 and by ARO MURI Award W911NF-14-0247. S.F. is supported by a Rutgers Center for Material Theory Distinguished Postdoctoral Fellowship. S.B.T. and J.C. recognize the support of the DOE Computational Science Graduate Fellowship (CSGF) under grant DE-FG02-97ER25308. This work used the Stampede2 supercomputer at the Texas Advanced Computing Center through allocation TG-DMR120073, which is part of the Extreme Science and Engineering Discovery Environment (XSEDE), supported by NSF Grant No. ACI-1548562. We also used the Odyssey cluster supported by the FAS Division of Science, Research Computing Group at Harvard University.
\end{acknowledgements}

\newpage
\begin{widetext}
\begin{appendices}

\section{\label{sec:level1}T-TMDC strain-dependent tight-binding parameters}

\input{multi-table.tex}

\end{appendices}
\end{widetext}

\clearpage
\bibliography{T_TMDC}

\end{document}

%% file: multi-table.tex
\begin{table}[h!]
  \centering
  \caption{T-type TMDC onsite strain terms in units of eV.}
  
\begin{tabular}{cccccccccc}
    & TiS$_2$ & TiSe$_2$ & TiTe$_2$ & NbS$_2$ & NbSe$_2$ & NbTe$_2$ & TaS$_2$ & TaSe$_2$ & TaTe$_2$ \\
    \hline
    \hline
	$\epsilon_{0}$  &  $-10.093$ &  $-9.195$ &  $-7.117$ &  $-9.023$ &  $-8.126$  &  $-6.178$  &  $-9.444$  &  $-8.582$  &  $-6.595$  \\ 
	$\epsilon_{1}$  &  $-10.030$ &  $-9.238$ &  $-7.324$ &  $-9.120$ &  $-8.322$  &  $-6.582$  &  $-9.532$  &  $-8.767$  &  $-7.041$  \\ 
	$\epsilon_{2}$  &  $-7.441$ &  $-6.809$ &  $-5.080$ &  $-6.523$ &  $-5.950$  &  $-4.413$  &  $-6.374$  &  $-5.906$  &  $-4.443$  \\ 
	$\epsilon_{3}$  &  $-7.072$ &  $-6.468$ &  $-4.776$ &  $-5.725$ &  $-5.256$  &  $-3.855$  &  $-5.375$  &  $-5.034$  &  $-3.706$  \\ 
	$\epsilon_{4}$  &  $-7.676$ &  $-6.998$ &  $-5.269$ &  $-6.803$ &  $-6.192$  &  $-4.694$  &  $-6.700$  &  $-6.197$  &  $-4.753$  \\ 
	$\epsilon_{5}$  &  $-0.442$ &  $-0.359$ &  $-0.304$ &  $-0.680$ &  $-0.548$  &  $-0.459$  &  $-0.779$  &  $-0.636$  &  $-0.511$  \\ 
	\hline
	$\alpha^{(0)}_{0}$  &  $-5.791$ &  $-5.989$ &  $-7.350$ &  $-5.618$ &  $-4.835$  &  $-6.604$  &  $-5.162$  &  $-5.132$  &  $-4.924$  \\ 
	$\alpha^{(0)}_{1}$  &  $-3.771$ &  $-3.816$ &  $-4.825$ &  $-4.046$ &  $-3.126$  &  $-4.564$  &  $-3.080$  &  $-2.837$  &  $-2.251$  \\ 
	$\alpha^{(0)}_{2}$  &  $-6.723$ &  $-6.992$ &  $-8.206$ &  $-5.086$ &  $-4.600$  &  $-6.592$  &  $-4.074$  &  $-4.568$  &  $-4.210$  \\ 
	$\alpha^{(0)}_{3}$  &  $-8.008$ &  $-8.008$ &  $-8.649$ &  $-6.718$ &  $-5.818$  &  $-6.767$  &  $-6.918$  &  $-6.695$  &  $-5.945$  \\ 
	$\alpha^{(0)}_{4}$  &  $-6.659$ &  $-6.855$ &  $-7.738$ &  $-4.898$ &  $-4.406$  &  $-5.744$  &  $-4.175$  &  $-4.368$  &  $-4.199$  \\ 
	$\alpha^{(0)}_{5}$  &  $ 0.408$ &  $ 0.301$ &  $ 0.196$ &  $ 0.245$ &  $ 0.137$  &  $ 0.130$  &  $-0.052$  &  $-0.052$  &  $-0.275$  \\ 
	\hline
	$\beta^{(0)}_{0}$  &  $-0.642$ &  $-0.604$ &  $-0.692$ &  $-0.477$ &  $-0.423$  &  $-0.589$  &  $-0.045$  &  $-0.033$  &  $-0.349$  \\ 
	$\beta^{(0)}_{1}$  &  $ 0.410$ &  $ 0.505$ &  $ 0.594$ &  $ 0.355$ &  $ 0.553$  &  $ 0.715$  &  $ 1.164$  &  $ 1.360$  &  $ 1.262$  \\ 
	$\beta^{(0)}_{2}$  &  $ 0.476$ &  $ 0.367$ &  $ 0.251$ &  $ 0.576$ &  $ 0.445$  &  $ 0.342$  &  $ 0.403$  &  $ 0.255$  &  $ 0.096$  \\ 
	$\beta^{(0)}_{3}$  &  $ 0.322$ &  $ 0.334$ &  $ 0.338$ &  $ 0.472$ &  $ 0.526$  &  $ 0.556$  &  $ 1.033$  &  $ 1.138$  &  $ 0.979$  \\ 
	$\beta^{(0)}_{4}$  &  $-0.554$ &  $-0.497$ &  $-0.334$ &  $-0.623$ &  $-0.450$  &  $-0.071$  &  $-0.609$  &  $-0.463$  &  $-0.032$  \\ 
	$\beta^{(0)}_{5}$  &  $-0.195$ &  $-0.167$ &  $-0.128$ &  $-0.227$ &  $-0.173$  &  $-0.202$  &  $-0.112$  &  $-0.036$  &  $ 0.009$  \\ 
	$\beta^{(0)}_{6}$  &  $-0.101$ &  $-0.131$ &  $-0.114$ &  $-0.041$ &  $-0.049$  &  $ 0.037$  &  $ 0.183$  &  $ 0.198$  &  $ 0.205$  \\ 
	\hline
	  \end{tabular}
\end{table}

\begin{table}[h!]
  \centering
  \caption{T-type TMDC 1NN isotropic strain terms in units of eV.}
  
  \label{tab:tmdc_1NN}
\begin{tabular}{cccccccccc}
    & TiS$_2$ & TiSe$_2$ & TiTe$_2$ & NbS$_2$ & NbSe$_2$ & NbTe$_2$ & TaS$_2$ & TaSe$_2$ & TaTe$_2$ \\
    \hline
    \hline
	$t^{(1)}_{0}$  &  $ 0.530$ &  $ 0.433$ &  $ 0.322$ &  $ 0.548$ &  $ 0.448$  &  $ 0.326$  &  $ 0.534$  &  $ 0.447$  &  $ 0.293$  \\ 
	$t^{(1)}_{1}$  &  $ 1.252$ &  $ 1.122$ &  $ 0.962$ &  $ 1.537$ &  $ 1.378$  &  $ 1.182$  &  $ 1.626$  &  $ 1.463$  &  $ 1.252$  \\ 
	$t^{(1)}_{2}$  &  $ 0.386$ &  $ 0.386$ &  $ 0.391$ &  $ 0.555$ &  $ 0.543$  &  $ 0.547$  &  $ 0.549$  &  $ 0.511$  &  $ 0.508$  \\ 
	$t^{(1)}_{3}$  &  $-0.698$ &  $-0.635$ &  $-0.565$ &  $-0.847$ &  $-0.772$  &  $-0.694$  &  $-0.925$  &  $-0.842$  &  $-0.766$  \\ 
	$t^{(1)}_{4}$  &  $-0.493$ &  $-0.441$ &  $-0.376$ &  $-0.597$ &  $-0.532$  &  $-0.437$  &  $-0.631$  &  $-0.559$  &  $-0.450$  \\ 
	$t^{(1)}_{5}$  &  $ 0.811$ &  $ 0.729$ &  $ 0.601$ &  $ 0.978$ &  $ 0.885$  &  $ 0.724$  &  $ 0.985$  &  $ 0.892$  &  $ 0.699$  \\ 
	$t^{(1)}_{6}$  &  $ 0.479$ &  $ 0.500$ &  $ 0.465$ &  $ 0.778$ &  $ 0.788$  &  $ 0.709$  &  $ 0.773$  &  $ 0.772$  &  $ 0.677$  \\ 
	$t^{(1)}_{7}$  &  $-0.574$ &  $-0.507$ &  $-0.423$ &  $-0.667$ &  $-0.586$  &  $-0.470$  &  $-0.756$  &  $-0.684$  &  $-0.575$  \\ 
	$t^{(1)}_{8}$  &  $ 0.220$ &  $ 0.201$ &  $ 0.194$ &  $ 0.145$ &  $ 0.135$  &  $ 0.128$  &  $ 0.190$  &  $ 0.186$  &  $ 0.179$  \\ 
	$t^{(1)}_{9}$  &  $-0.112$ &  $-0.102$ &  $-0.065$ &  $-0.127$ &  $-0.110$  &  $-0.052$  &  $-0.175$  &  $-0.169$  &  $-0.100$  \\ 
	$t^{(1)}_{10}$  &  $ 0.430$ &  $ 0.454$ &  $ 0.455$ &  $ 0.356$ &  $ 0.395$  &  $ 0.409$  &  $ 0.390$  &  $ 0.432$  &  $ 0.428$  \\ 
	$t^{(1)}_{11}$  &  $-0.363$ &  $-0.349$ &  $-0.292$ &  $-0.262$ &  $-0.262$  &  $-0.208$  &  $-0.285$  &  $-0.295$  &  $-0.228$  \\ 
	\hline
	$\alpha^{(1)}_{0}$  &  $ 0.667$ &  $ 0.634$ &  $ 0.568$ &  $ 0.914$ &  $ 0.850$  &  $ 0.669$  &  $ 1.520$  &  $ 1.341$  &  $ 1.695$  \\ 
	$\alpha^{(1)}_{1}$  &  $-0.659$ &  $-0.454$ &  $-0.257$ &  $-0.380$ &  $-0.153$  &  $-0.032$  &  $-0.362$  &  $-0.221$  &  $-0.049$  \\ 
	$\alpha^{(1)}_{2}$  &  $-1.169$ &  $-1.002$ &  $-0.888$ &  $-1.274$ &  $-1.085$  &  $-0.945$  &  $-1.187$  &  $-1.033$  &  $-0.779$  \\ 
	$\alpha^{(1)}_{3}$  &  $ 0.750$ &  $ 0.624$ &  $ 0.516$ &  $ 0.538$ &  $ 0.396$  &  $ 0.302$  &  $ 0.927$  &  $ 0.785$  &  $ 0.709$  \\ 
	$\alpha^{(1)}_{4}$  &  $ 0.363$ &  $ 0.271$ &  $ 0.095$ &  $ 0.203$ &  $ 0.097$  &  $-0.121$  &  $-0.168$  &  $-0.189$  &  $-0.744$  \\ 
	$\alpha^{(1)}_{5}$  &  $-0.407$ &  $-0.289$ &  $-0.124$ &  $-0.175$ &  $-0.065$  &  $ 0.015$  &  $ 0.183$  &  $ 0.220$  &  $ 0.621$  \\ 
	$\alpha^{(1)}_{6}$  &  $-1.381$ &  $-1.214$ &  $-0.864$ &  $-1.750$ &  $-1.506$  &  $-1.052$  &  $-1.791$  &  $-1.513$  &  $-1.274$  \\ 
	$\alpha^{(1)}_{7}$  &  $ 0.281$ &  $ 0.187$ &  $ 0.025$ &  $-0.116$ &  $-0.210$  &  $-0.363$  &  $ 0.105$  &  $ 0.056$  &  $-0.076$  \\ 
	$\alpha^{(1)}_{8}$  &  $ 0.550$ &  $ 0.526$ &  $ 0.523$ &  $ 0.395$ &  $ 0.418$  &  $ 0.438$  &  $ 0.511$  &  $ 0.526$  &  $ 0.551$  \\ 
	$\alpha^{(1)}_{9}$  &  $-0.435$ &  $-0.420$ &  $-0.425$ &  $-0.413$ &  $-0.403$  &  $-0.388$  &  $-0.533$  &  $-0.515$  &  $-0.462$  \\ 
	$\alpha^{(1)}_{10}$  &  $-0.001$ &  $ 0.040$ &  $ 0.083$ &  $ 0.037$ &  $ 0.047$  &  $ 0.074$  &  $ 0.134$  &  $ 0.141$  &  $ 0.232$  \\ 
	$\alpha^{(1)}_{11}$  &  $-0.625$ &  $-0.617$ &  $-0.639$ &  $-0.539$ &  $-0.538$  &  $-0.551$  &  $-0.747$  &  $-0.746$  &  $-0.742$  \\ 
	\hline
	  \end{tabular}
\end{table}

\begin{table}[h!]
  \centering
  \caption{T-type TMDC 1NN anisotropic strain terms in units of eV.}
  
  \label{tab:tmdc_1NN_aniso}
\begin{tabular}{cccccccccc}
	    & TiS$_2$ & TiSe$_2$ & TiTe$_2$ & NbS$_2$ & NbSe$_2$ & NbTe$_2$ & TaS$_2$ & TaSe$_2$ & TaTe$_2$ \\
    \hline
    \hline
	$\beta^{(1)}_{0}$  &  $ 0.129$ &  $ 0.193$ &  $ 0.208$ &  $ 0.423$ &  $ 0.482$  &  $ 0.445$  &  $ 1.069$  &  $ 1.102$  &  $ 0.805$  \\ 
	$\beta^{(1)}_{1}$  &  $-0.320$ &  $-0.115$ &  $ 0.019$ &  $ 0.071$ &  $ 0.282$  &  $ 0.294$  &  $ 1.094$  &  $ 1.197$  &  $ 0.952$  \\ 
	$\beta^{(1)}_{2}$  &  $-1.185$ &  $-1.115$ &  $-0.994$ &  $-1.398$ &  $-1.283$  &  $-1.125$  &  $-1.340$  &  $-1.191$  &  $-1.119$  \\ 
	$\beta^{(1)}_{3}$  &  $ 1.026$ &  $ 0.883$ &  $ 0.751$ &  $ 1.056$ &  $ 0.925$  &  $ 0.818$  &  $ 1.150$  &  $ 1.002$  &  $ 0.889$  \\ 
	$\beta^{(1)}_{4}$  &  $ 0.670$ &  $ 0.594$ &  $ 0.467$ &  $ 0.849$ &  $ 0.724$  &  $ 0.532$  &  $ 1.034$  &  $ 0.923$  &  $ 0.665$  \\ 
	$\beta^{(1)}_{5}$  &  $-0.736$ &  $-0.573$ &  $-0.326$ &  $-0.673$ &  $-0.441$  &  $-0.140$  &  $-0.336$  &  $-0.116$  &  $ 0.091$  \\ 
	$\beta^{(1)}_{6}$  &  $-0.400$ &  $-0.327$ &  $-0.242$ &  $-0.505$ &  $-0.335$  &  $-0.234$  &  $ 0.125$  &  $ 0.287$  &  $ 0.278$  \\ 
	$\beta^{(1)}_{7}$  &  $ 0.530$ &  $ 0.409$ &  $ 0.309$ &  $ 0.470$ &  $ 0.350$  &  $ 0.264$  &  $ 0.722$  &  $ 0.662$  &  $ 0.552$  \\ 
	$\beta^{(1)}_{8}$  &  $ 1.283$ &  $ 1.139$ &  $ 0.986$ &  $ 1.538$ &  $ 1.413$  &  $ 1.168$  &  $ 1.674$  &  $ 1.460$  &  $ 1.094$  \\ 
	$\beta^{(1)}_{9}$  &  $-0.263$ &  $-0.376$ &  $-0.430$ &  $-0.605$ &  $-0.762$  &  $-0.769$  &  $-1.513$  &  $-1.632$  &  $-1.380$  \\ 
	$\beta^{(1)}_{10}$  &  $ 0.792$ &  $ 0.672$ &  $ 0.515$ &  $ 0.867$ &  $ 0.742$  &  $ 0.551$  &  $ 0.807$  &  $ 0.670$  &  $ 0.492$  \\ 
	$\beta^{(1)}_{11}$  &  $ 0.530$ &  $ 0.444$ &  $ 0.272$ &  $ 0.626$ &  $ 0.512$  &  $ 0.275$  &  $ 0.424$  &  $ 0.283$  &  $ 0.083$  \\ 
	$\beta^{(1)}_{12}$  &  $ 0.119$ &  $ 0.125$ &  $ 0.132$ &  $ 0.270$ &  $ 0.237$  &  $ 0.256$  &  $ 0.219$  &  $ 0.180$  &  $ 0.143$  \\ 
	$\beta^{(1)}_{13}$  &  $ 0.752$ &  $ 0.707$ &  $ 0.591$ &  $ 1.079$ &  $ 1.035$  &  $ 0.809$  &  $ 1.028$  &  $ 0.926$  &  $ 0.648$  \\ 
	$\beta^{(1)}_{14}$  &  $-0.138$ &  $-0.209$ &  $-0.172$ &  $-0.349$ &  $-0.449$  &  $-0.296$  &  $-0.965$  &  $-1.137$  &  $-0.733$  \\ 
	$\beta^{(1)}_{15}$  &  $ 0.189$ &  $ 0.221$ &  $ 0.196$ &  $ 0.289$ &  $ 0.327$  &  $ 0.223$  &  $ 0.304$  &  $ 0.336$  &  $ 0.238$  \\ 
	$\beta^{(1)}_{16}$  &  $ 0.086$ &  $ 0.032$ &  $ 0.011$ &  $-0.105$ &  $-0.175$  &  $-0.112$  &  $-0.360$  &  $-0.430$  &  $-0.300$  \\ 
	$\beta^{(1)}_{17}$  &  $-0.371$ &  $-0.279$ &  $-0.221$ &  $-0.141$ &  $-0.028$  &  $-0.079$  &  $-0.236$  &  $-0.141$  &  $-0.212$  \\ 
	$\beta^{(1)}_{18}$  &  $ 0.028$ &  $-0.064$ &  $-0.161$ &  $-0.230$ &  $-0.327$  &  $-0.375$  &  $-0.336$  &  $-0.398$  &  $-0.407$  \\ 
	$\beta^{(1)}_{19}$  &  $ 0.080$ &  $ 0.074$ &  $ 0.033$ &  $ 0.107$ &  $ 0.081$  &  $ 0.031$  &  $ 0.192$  &  $ 0.169$  &  $ 0.076$  \\ 
	$\beta^{(1)}_{20}$  &  $-0.313$ &  $-0.287$ &  $-0.264$ &  $-0.155$ &  $-0.133$  &  $-0.108$  &  $-0.107$  &  $-0.092$  &  $-0.028$  \\ 
	\hline
	  \end{tabular}
\end{table}

\begin{table}[h!]
  \centering
  \caption{T-type TMDC 2NN isotropic strain terms in units of eV.}
  
  \label{tab:tmdc_2NN}
\begin{tabular}{cccccccccc}
	    & TiS$_2$ & TiSe$_2$ & TiTe$_2$ & NbS$_2$ & NbSe$_2$ & NbTe$_2$ & TaS$_2$ & TaSe$_2$ & TaTe$_2$ \\
    \hline
    \hline
	$t^{(2)}_{0}$  &  $-0.067$ &  $-0.072$ &  $-0.040$ &  $-0.087$ &  $-0.089$  &  $-0.039$  &  $-0.077$  &  $-0.081$  &  $-0.026$  \\ 
	$t^{(2)}_{1}$  &  $ 0.701$ &  $ 0.782$ &  $ 0.883$ &  $ 0.732$ &  $ 0.831$  &  $ 0.982$  &  $ 0.762$  &  $ 0.842$  &  $ 0.989$  \\ 
	$t^{(2)}_{2}$  &  $-0.113$ &  $-0.137$ &  $-0.162$ &  $-0.159$ &  $-0.187$  &  $-0.219$  &  $-0.153$  &  $-0.176$  &  $-0.200$  \\ 
	$t^{(2)}_{3}$  &  $-0.050$ &  $-0.051$ &  $-0.051$ &  $-0.047$ &  $-0.048$  &  $-0.040$  &  $-0.076$  &  $-0.080$  &  $-0.071$  \\ 
	$t^{(2)}_{4}$  &  $ 0.015$ &  $ 0.019$ &  $ 0.030$ &  $ 0.039$ &  $ 0.040$  &  $ 0.046$  &  $ 0.062$  &  $ 0.065$  &  $ 0.072$  \\ 
	$t^{(2)}_{5}$  &  $-0.018$ &  $-0.034$ &  $-0.034$ &  $-0.036$ &  $-0.049$  &  $-0.037$  &  $-0.067$  &  $-0.085$  &  $-0.066$  \\ 
	$t^{(2)}_{6}$  &  $ 0.032$ &  $ 0.056$ &  $ 0.072$ &  $ 0.140$ &  $ 0.169$  &  $ 0.182$  &  $ 0.188$  &  $ 0.199$  &  $ 0.229$  \\ 
	$t^{(2)}_{7}$  &  $-0.125$ &  $-0.091$ &  $-0.074$ &  $-0.188$ &  $-0.121$  &  $-0.085$  &  $-0.205$  &  $-0.140$  &  $-0.079$  \\ 
	$t^{(2)}_{8}$  &  $-0.214$ &  $-0.201$ &  $-0.188$ &  $-0.467$ &  $-0.432$  &  $-0.398$  &  $-0.547$  &  $-0.486$  &  $-0.453$  \\ 
	$t^{(2)}_{9}$  &  $-0.021$ &  $-0.016$ &  $-0.004$ &  $-0.006$ &  $-0.005$  &  $ 0.011$  &  $-0.028$  &  $-0.028$  &  $-0.007$  \\ 
	$t^{(2)}_{10}$  &  $-0.103$ &  $-0.108$ &  $-0.113$ &  $-0.267$ &  $-0.263$  &  $-0.254$  &  $-0.313$  &  $-0.297$  &  $-0.304$  \\ 
	$t^{(2)}_{11}$  &  $ 0.210$ &  $ 0.166$ &  $ 0.141$ &  $ 0.339$ &  $ 0.265$  &  $ 0.225$  &  $ 0.365$  &  $ 0.282$  &  $ 0.222$  \\ 
	$t^{(2)}_{12}$  &  $ 0.114$ &  $ 0.088$ &  $ 0.070$ &  $ 0.158$ &  $ 0.117$  &  $ 0.082$  &  $ 0.162$  &  $ 0.121$  &  $ 0.074$  \\ 
	$t^{(2)}_{13}$  &  $-0.196$ &  $-0.182$ &  $-0.170$ &  $-0.387$ &  $-0.351$  &  $-0.316$  &  $-0.445$  &  $-0.395$  &  $-0.368$  \\ 
	$t^{(2)}_{14}$  &  $ 0.094$ &  $ 0.077$ &  $ 0.058$ &  $ 0.126$ &  $ 0.096$  &  $ 0.053$  &  $ 0.147$  &  $ 0.116$  &  $ 0.063$  \\ 
	\hline
	$\alpha^{(2)}_{0}$  &  $ 0.059$ &  $ 0.036$ &  $-0.115$ &  $ 0.039$ &  $ 0.012$  &  $-0.168$  &  $-0.046$  &  $-0.076$  &  $-0.197$  \\ 
	$\alpha^{(2)}_{1}$  &  $-1.344$ &  $-1.493$ &  $-1.681$ &  $-1.329$ &  $-1.523$  &  $-1.822$  &  $-1.256$  &  $-1.403$  &  $-1.744$  \\ 
	$\alpha^{(2)}_{2}$  &  $ 0.408$ &  $ 0.464$ &  $ 0.548$ &  $ 0.493$ &  $ 0.557$  &  $ 0.672$  &  $ 0.396$  &  $ 0.438$  &  $ 0.534$  \\ 
	$\alpha^{(2)}_{3}$  &  $-0.104$ &  $-0.102$ &  $-0.118$ &  $-0.132$ &  $-0.130$  &  $-0.135$  &  $-0.191$  &  $-0.192$  &  $-0.138$  \\ 
	$\alpha^{(2)}_{4}$  &  $-0.030$ &  $-0.023$ &  $-0.014$ &  $-0.077$ &  $-0.053$  &  $-0.005$  &  $-0.039$  &  $-0.030$  &  $ 0.028$  \\ 
	$\alpha^{(2)}_{5}$  &  $ 0.020$ &  $ 0.018$ &  $-0.031$ &  $-0.104$ &  $-0.088$  &  $-0.129$  &  $-0.114$  &  $-0.103$  &  $-0.176$  \\ 
	$\alpha^{(2)}_{6}$  &  $-0.581$ &  $-0.576$ &  $-0.536$ &  $-1.242$ &  $-1.177$  &  $-1.038$  &  $-1.608$  &  $-1.394$  &  $-1.389$  \\ 
	$\alpha^{(2)}_{7}$  &  $ 0.128$ &  $-0.006$ &  $-0.135$ &  $-0.258$ &  $-0.452$  &  $-0.529$  &  $-0.146$  &  $-0.296$  &  $-0.264$  \\ 
	$\alpha^{(2)}_{8}$  &  $ 0.944$ &  $ 0.835$ &  $ 0.722$ &  $ 1.608$ &  $ 1.418$  &  $ 1.188$  &  $ 1.885$  &  $ 1.613$  &  $ 1.387$  \\ 
	$\alpha^{(2)}_{9}$  &  $-0.202$ &  $-0.154$ &  $-0.119$ &  $-0.357$ &  $-0.248$  &  $-0.187$  &  $-0.422$  &  $-0.276$  &  $-0.291$  \\ 
	$\alpha^{(2)}_{10}$  &  $ 0.677$ &  $ 0.638$ &  $ 0.568$ &  $ 1.200$ &  $ 1.099$  &  $ 0.863$  &  $ 1.480$  &  $ 1.335$  &  $ 1.255$  \\ 
	$\alpha^{(2)}_{11}$  &  $-0.383$ &  $-0.284$ &  $-0.223$ &  $-0.292$ &  $-0.205$  &  $-0.192$  &  $-0.140$  &  $-0.125$  &  $-0.061$  \\ 
	$\alpha^{(2)}_{12}$  &  $-0.069$ &  $-0.024$ &  $ 0.031$ &  $ 0.127$ &  $ 0.163$  &  $ 0.202$  &  $ 0.377$  &  $ 0.341$  &  $ 0.604$  \\ 
	$\alpha^{(2)}_{13}$  &  $ 0.727$ &  $ 0.622$ &  $ 0.503$ &  $ 1.028$ &  $ 0.851$  &  $ 0.599$  &  $ 1.260$  &  $ 1.074$  &  $ 0.882$  \\ 
	$\alpha^{(2)}_{14}$  &  $-0.047$ &  $-0.000$ &  $ 0.089$ &  $ 0.227$ &  $ 0.253$  &  $ 0.351$  &  $ 0.389$  &  $ 0.372$  &  $ 0.649$  \\ 
	\hline
	  \end{tabular}
\end{table}

\begin{table}[h!]
  \centering
  \caption{T-type TMDC 2NN anisotropic strain terms in units of eV.}
  
  \label{tab:tmdc_2NN_aniso}
\begin{tabular}{cccccccccc}	
  & TiS$_2$ & TiSe$_2$ & TiTe$_2$ & NbS$_2$ & NbSe$_2$ & NbTe$_2$ & TaS$_2$ & TaSe$_2$ & TaTe$_2$ \\
    \hline
    \hline
	$\beta^{(2)}_{0}$  &  $-0.516$ &  $-0.570$ &  $-0.718$ &  $-0.544$ &  $-0.616$  &  $-0.839$  &  $-0.541$  &  $-0.595$  &  $-0.798$  \\ 
	$\beta^{(2)}_{1}$  &  $ 1.656$ &  $ 1.835$ &  $ 2.106$ &  $ 1.734$ &  $ 1.965$  &  $ 2.356$  &  $ 1.854$  &  $ 2.025$  &  $ 2.371$  \\ 
	$\beta^{(2)}_{2}$  &  $-0.439$ &  $-0.502$ &  $-0.573$ &  $-0.574$ &  $-0.655$  &  $-0.721$  &  $-0.526$  &  $-0.584$  &  $-0.592$  \\ 
	$\beta^{(2)}_{3}$  &  $-0.100$ &  $-0.121$ &  $-0.121$ &  $-0.152$ &  $-0.191$  &  $-0.151$  &  $-0.286$  &  $-0.320$  &  $-0.246$  \\ 
	$\beta^{(2)}_{4}$  &  $ 0.085$ &  $ 0.088$ &  $ 0.081$ &  $ 0.170$ &  $ 0.172$  &  $ 0.113$  &  $ 0.149$  &  $ 0.133$  &  $ 0.101$  \\ 
	$\beta^{(2)}_{5}$  &  $-0.024$ &  $-0.063$ &  $-0.035$ &  $-0.057$ &  $-0.119$  &  $-0.052$  &  $-0.175$  &  $-0.243$  &  $-0.150$  \\ 
	$\beta^{(2)}_{6}$  &  $ 0.475$ &  $ 0.524$ &  $ 0.610$ &  $ 0.489$ &  $ 0.561$  &  $ 0.705$  &  $ 0.521$  &  $ 0.577$  &  $ 0.731$  \\ 
	$\beta^{(2)}_{7}$  &  $ 0.034$ &  $ 0.023$ &  $ 0.027$ &  $ 0.036$ &  $ 0.020$  &  $ 0.027$  &  $ 0.137$  &  $ 0.130$  &  $ 0.129$  \\ 
	$\beta^{(2)}_{8}$  &  $-0.010$ &  $-0.021$ &  $-0.039$ &  $-0.008$ &  $-0.038$  &  $-0.074$  &  $-0.054$  &  $-0.088$  &  $-0.140$  \\ 
	$\beta^{(2)}_{9}$  &  $ 0.315$ &  $ 0.332$ &  $ 0.341$ &  $ 0.820$ &  $ 0.780$  &  $ 0.727$  &  $ 1.128$  &  $ 1.031$  &  $ 1.000$  \\ 
	$\beta^{(2)}_{10}$  &  $ 0.160$ &  $ 0.175$ &  $ 0.158$ &  $ 0.601$ &  $ 0.566$  &  $ 0.482$  &  $ 0.546$  &  $ 0.441$  &  $ 0.472$  \\ 
	$\beta^{(2)}_{11}$  &  $-0.922$ &  $-0.758$ &  $-0.622$ &  $-1.406$ &  $-1.140$  &  $-0.947$  &  $-1.544$  &  $-1.260$  &  $-1.187$  \\ 
	$\beta^{(2)}_{12}$  &  $ 0.019$ &  $ 0.002$ &  $-0.005$ &  $-0.066$ &  $-0.101$  &  $-0.120$  &  $-0.382$  &  $-0.410$  &  $-0.413$  \\ 
	$\beta^{(2)}_{13}$  &  $-0.233$ &  $-0.205$ &  $-0.152$ &  $-0.376$ &  $-0.340$  &  $-0.224$  &  $-0.446$  &  $-0.431$  &  $-0.381$  \\ 
	$\beta^{(2)}_{14}$  &  $ 0.464$ &  $ 0.357$ &  $ 0.269$ &  $ 0.516$ &  $ 0.389$  &  $ 0.272$  &  $ 0.474$  &  $ 0.363$  &  $ 0.182$  \\ 
	$\beta^{(2)}_{15}$  &  $ 0.127$ &  $ 0.100$ &  $ 0.037$ &  $ 0.121$ &  $ 0.103$  &  $-0.020$  &  $ 0.203$  &  $ 0.218$  &  $-0.028$  \\ 
	$\beta^{(2)}_{16}$  &  $-0.552$ &  $-0.470$ &  $-0.378$ &  $-0.868$ &  $-0.716$  &  $-0.540$  &  $-0.958$  &  $-0.812$  &  $-0.684$  \\ 
	$\beta^{(2)}_{17}$  &  $ 0.067$ &  $ 0.068$ &  $ 0.049$ &  $ 0.152$ &  $ 0.154$  &  $ 0.075$  &  $ 0.350$  &  $ 0.372$  &  $ 0.262$  \\ 
	$\beta^{(2)}_{18}$  &  $ 0.216$ &  $ 0.221$ &  $ 0.198$ &  $ 0.545$ &  $ 0.544$  &  $ 0.478$  &  $ 0.641$  &  $ 0.612$  &  $ 0.516$  \\ 
	$\beta^{(2)}_{19}$  &  $-0.130$ &  $-0.120$ &  $-0.145$ &  $-0.235$ &  $-0.220$  &  $-0.276$  &  $-0.312$  &  $-0.272$  &  $-0.339$  \\ 
	$\beta^{(2)}_{20}$  &  $ 0.167$ &  $ 0.149$ &  $ 0.117$ &  $ 0.392$ &  $ 0.365$  &  $ 0.281$  &  $ 0.394$  &  $ 0.339$  &  $ 0.248$  \\ 
	$\beta^{(2)}_{21}$  &  $-0.006$ &  $-0.013$ &  $-0.016$ &  $-0.061$ &  $-0.088$  &  $-0.086$  &  $-0.141$  &  $-0.144$  &  $-0.137$  \\ 
	$\beta^{(2)}_{22}$  &  $-0.076$ &  $-0.048$ &  $-0.030$ &  $-0.027$ &  $ 0.008$  &  $ 0.030$  &  $ 0.153$  &  $ 0.199$  &  $ 0.182$  \\ 
	$\beta^{(2)}_{23}$  &  $-0.161$ &  $-0.151$ &  $-0.111$ &  $-0.390$ &  $-0.368$  &  $-0.255$  &  $-0.451$  &  $-0.399$  &  $-0.220$  \\ 
	\hline
	  \end{tabular}
\end{table}

\begin{table}[h!]
  \centering
  \caption{T-type TMDC 3NN isotropic strain terms in units of eV.}
  
  \label{tab:tmdc_3NN}
\begin{tabular}{cccccccccc}
& TiS$_2$ & TiSe$_2$ & TiTe$_2$ & NbS$_2$ & NbSe$_2$ & NbTe$_2$ & TaS$_2$ & TaSe$_2$ & TaTe$_2$ \\
    \hline
    \hline
	$t^{(3)}_{0}$  &  $-0.055$ &  $-0.054$ &  $-0.058$ &  $-0.077$ &  $-0.071$  &  $-0.075$  &  $-0.087$  &  $-0.078$  &  $-0.082$  \\ 
	$t^{(3)}_{1}$  &  $ 0.018$ &  $ 0.008$ &  $-0.014$ &  $-0.005$ &  $-0.015$  &  $-0.052$  &  $-0.016$  &  $-0.024$  &  $-0.067$  \\ 
	$t^{(3)}_{2}$  &  $ 0.037$ &  $ 0.051$ &  $ 0.063$ &  $ 0.093$ &  $ 0.107$  &  $ 0.120$  &  $ 0.095$  &  $ 0.102$  &  $ 0.109$  \\ 
	$t^{(3)}_{3}$  &  $-0.057$ &  $-0.052$ &  $-0.064$ &  $-0.048$ &  $-0.042$  &  $-0.061$  &  $-0.086$  &  $-0.079$  &  $-0.098$  \\ 
	$t^{(3)}_{4}$  &  $ 0.022$ &  $ 0.020$ &  $ 0.048$ &  $ 0.046$ &  $ 0.041$  &  $ 0.083$  &  $ 0.088$  &  $ 0.076$  &  $ 0.131$  \\ 
	$t^{(3)}_{5}$  &  $ 0.011$ &  $ 0.027$ &  $ 0.034$ &  $ 0.049$ &  $ 0.070$  &  $ 0.085$  &  $ 0.037$  &  $ 0.054$  &  $ 0.073$  \\ 
	$t^{(3)}_{6}$  &  $-0.060$ &  $-0.057$ &  $-0.083$ &  $-0.091$ &  $-0.089$  &  $-0.130$  &  $-0.125$  &  $-0.115$  &  $-0.163$  \\ 
	$t^{(3)}_{7}$  &  $ 0.014$ &  $ 0.008$ &  $ 0.000$ &  $-0.016$ &  $-0.023$  &  $-0.037$  &  $-0.013$  &  $-0.014$  &  $-0.025$  \\ 
	$t^{(3)}_{8}$  &  $-0.040$ &  $-0.041$ &  $-0.051$ &  $-0.044$ &  $-0.046$  &  $-0.055$  &  $-0.009$  &  $-0.011$  &  $-0.020$  \\ 
	$t^{(3)}_{9}$  &  $-0.086$ &  $-0.080$ &  $-0.068$ &  $-0.131$ &  $-0.119$  &  $-0.098$  &  $-0.153$  &  $-0.144$  &  $-0.123$  \\ 
	$t^{(3)}_{10}$  &  $-0.072$ &  $-0.073$ &  $-0.065$ &  $-0.089$ &  $-0.089$  &  $-0.061$  &  $-0.054$  &  $-0.050$  &  $-0.005$  \\ 
	$t^{(3)}_{11}$  &  $ 0.028$ &  $ 0.024$ &  $ 0.015$ &  $ 0.065$ &  $ 0.057$  &  $ 0.054$  &  $ 0.119$  &  $ 0.112$  &  $ 0.116$  \\ 
	\hline
	$\alpha^{(3)}_{0}$  &  $ 0.314$ &  $ 0.266$ &  $ 0.233$ &  $ 0.242$ &  $ 0.194$  &  $ 0.169$  &  $ 0.334$  &  $ 0.276$  &  $ 0.119$  \\ 
	$\alpha^{(3)}_{1}$  &  $ 0.122$ &  $ 0.118$ &  $ 0.172$ &  $ 0.259$ &  $ 0.227$  &  $ 0.290$  &  $ 0.413$  &  $ 0.362$  &  $ 0.373$  \\ 
	$\alpha^{(3)}_{2}$  &  $-0.411$ &  $-0.429$ &  $-0.426$ &  $-0.520$ &  $-0.528$  &  $-0.492$  &  $-0.607$  &  $-0.614$  &  $-0.491$  \\ 
	$\alpha^{(3)}_{3}$  &  $ 0.024$ &  $ 0.009$ &  $-0.006$ &  $-0.097$ &  $-0.101$  &  $-0.117$  &  $-0.171$  &  $-0.139$  &  $-0.257$  \\ 
	$\alpha^{(3)}_{4}$  &  $-0.220$ &  $-0.201$ &  $-0.234$ &  $-0.301$ &  $-0.245$  &  $-0.216$  &  $-0.477$  &  $-0.447$  &  $-0.321$  \\ 
	$\alpha^{(3)}_{5}$  &  $-0.273$ &  $-0.317$ &  $-0.366$ &  $-0.423$ &  $-0.484$  &  $-0.571$  &  $-0.494$  &  $-0.525$  &  $-0.681$  \\ 
	$\alpha^{(3)}_{6}$  &  $ 0.323$ &  $ 0.306$ &  $ 0.345$ &  $ 0.377$ &  $ 0.352$  &  $ 0.313$  &  $ 0.528$  &  $ 0.511$  &  $ 0.414$  \\ 
	$\alpha^{(3)}_{7}$  &  $ 0.184$ &  $ 0.211$ &  $ 0.246$ &  $ 0.293$ &  $ 0.330$  &  $ 0.363$  &  $ 0.368$  &  $ 0.400$  &  $ 0.389$  \\ 
	$\alpha^{(3)}_{8}$  &  $ 0.054$ &  $ 0.051$ &  $ 0.068$ &  $ 0.088$ &  $ 0.094$  &  $ 0.112$  &  $ 0.073$  &  $ 0.115$  &  $ 0.086$  \\ 
	$\alpha^{(3)}_{9}$  &  $ 0.085$ &  $ 0.052$ &  $ 0.012$ &  $ 0.014$ &  $-0.034$  &  $-0.082$  &  $-0.006$  &  $-0.037$  &  $-0.095$  \\ 
	$\alpha^{(3)}_{10}$  &  $ 0.088$ &  $ 0.097$ &  $ 0.053$ &  $ 0.005$ &  $ 0.023$  &  $-0.018$  &  $-0.228$  &  $-0.232$  &  $-0.233$  \\ 
	$\alpha^{(3)}_{11}$  &  $-0.044$ &  $-0.022$ &  $-0.009$ &  $-0.065$ &  $-0.014$  &  $ 0.033$  &  $-0.220$  &  $-0.184$  &  $-0.118$  \\ 
	\hline
  \end{tabular}
\end{table}

\begin{table}[h!]
  \centering
  \caption{T-type TMDC 3NN anisotropic strain terms in units of eV.}
  
  \label{tab:tmdc_3NN_aniso}
\begin{tabular}{cccccccccc}	
& TiS$_2$ & TiSe$_2$ & TiTe$_2$ & NbS$_2$ & NbSe$_2$ & NbTe$_2$ & TaS$_2$ & TaSe$_2$ & TaTe$_2$ \\
    \hline
    \hline
	$\beta^{(3)}_{0}$  &  $ 0.524$ &  $ 0.515$ &  $ 0.559$ &  $ 0.625$ &  $ 0.593$  &  $ 0.674$  &  $ 0.709$  &  $ 0.668$  &  $ 0.708$  \\ 
	$\beta^{(3)}_{1}$  &  $-0.461$ &  $-0.424$ &  $-0.318$ &  $-0.565$ &  $-0.518$  &  $-0.333$  &  $-0.473$  &  $-0.427$  &  $-0.236$  \\ 
	$\beta^{(3)}_{2}$  &  $-0.096$ &  $-0.105$ &  $-0.095$ &  $-0.098$ &  $-0.102$  &  $-0.067$  &  $-0.111$  &  $-0.117$  &  $-0.074$  \\ 
	$\beta^{(3)}_{3}$  &  $-0.012$ &  $-0.015$ &  $-0.010$ &  $-0.085$ &  $-0.079$  &  $-0.061$  &  $-0.193$  &  $-0.184$  &  $-0.121$  \\ 
	$\beta^{(3)}_{4}$  &  $-0.135$ &  $-0.125$ &  $-0.136$ &  $-0.191$ &  $-0.175$  &  $-0.186$  &  $-0.092$  &  $-0.049$  &  $-0.096$  \\ 
	$\beta^{(3)}_{5}$  &  $-0.004$ &  $-0.054$ &  $-0.079$ &  $-0.051$ &  $-0.113$  &  $-0.159$  &  $ 0.006$  &  $-0.063$  &  $-0.138$  \\ 
	$\beta^{(3)}_{6}$  &  $ 0.240$ &  $ 0.231$ &  $ 0.238$ &  $ 0.327$ &  $ 0.314$  &  $ 0.333$  &  $ 0.489$  &  $ 0.438$  &  $ 0.457$  \\ 
	$\beta^{(3)}_{7}$  &  $ 0.110$ &  $ 0.133$ &  $ 0.178$ &  $ 0.234$ &  $ 0.256$  &  $ 0.305$  &  $ 0.306$  &  $ 0.318$  &  $ 0.376$  \\ 
	$\beta^{(3)}_{8}$  &  $ 0.158$ &  $ 0.138$ &  $ 0.160$ &  $ 0.223$ &  $ 0.209$  &  $ 0.284$  &  $ 0.249$  &  $ 0.228$  &  $ 0.271$  \\ 
	$\beta^{(3)}_{9}$  &  $ 0.073$ &  $ 0.076$ &  $ 0.005$ &  $ 0.178$ &  $ 0.144$  &  $-0.035$  &  $ 0.210$  &  $ 0.174$  &  $-0.053$  \\ 
	$\beta^{(3)}_{10}$  &  $-0.227$ &  $-0.226$ &  $-0.258$ &  $-0.330$ &  $-0.311$  &  $-0.339$  &  $-0.452$  &  $-0.426$  &  $-0.423$  \\ 
	$\beta^{(3)}_{11}$  &  $ 0.115$ &  $ 0.108$ &  $ 0.093$ &  $ 0.118$ &  $ 0.118$  &  $ 0.096$  &  $ 0.024$  &  $ 0.010$  &  $ 0.004$  \\ 
	$\beta^{(3)}_{12}$  &  $ 0.148$ &  $ 0.159$ &  $ 0.204$ &  $ 0.189$ &  $ 0.202$  &  $ 0.270$  &  $ 0.227$  &  $ 0.232$  &  $ 0.293$  \\ 
	$\beta^{(3)}_{13}$  &  $-0.120$ &  $-0.106$ &  $-0.131$ &  $-0.150$ &  $-0.132$  &  $-0.190$  &  $-0.290$  &  $-0.270$  &  $-0.261$  \\ 
	$\beta^{(3)}_{14}$  &  $-0.015$ &  $-0.018$ &  $ 0.010$ &  $ 0.003$ &  $ 0.004$  &  $ 0.078$  &  $ 0.138$  &  $ 0.159$  &  $ 0.174$  \\ 
	$\beta^{(3)}_{15}$  &  $-0.351$ &  $-0.350$ &  $-0.288$ &  $-0.560$ &  $-0.574$  &  $-0.419$  &  $-1.296$  &  $-1.281$  &  $-0.934$  \\ 
	$\beta^{(3)}_{16}$  &  $ 0.177$ &  $ 0.155$ &  $ 0.131$ &  $ 0.242$ &  $ 0.211$  &  $ 0.174$  &  $ 0.305$  &  $ 0.278$  &  $ 0.250$  \\ 
	$\beta^{(3)}_{17}$  &  $-0.036$ &  $-0.075$ &  $-0.105$ &  $-0.116$ &  $-0.189$  &  $-0.235$  &  $-0.336$  &  $-0.401$  &  $-0.365$  \\ 
	$\beta^{(3)}_{18}$  &  $-0.306$ &  $-0.308$ &  $-0.266$ &  $-0.508$ &  $-0.538$  &  $-0.461$  &  $-0.956$  &  $-0.979$  &  $-0.791$  \\ 
	$\beta^{(3)}_{19}$  &  $ 0.041$ &  $ 0.042$ &  $ 0.042$ &  $ 0.095$ &  $ 0.110$  &  $ 0.104$  &  $ 0.162$  &  $ 0.179$  &  $ 0.185$  \\ 
	$\beta^{(3)}_{20}$  &  $-0.030$ &  $-0.044$ &  $-0.026$ &  $-0.034$ &  $-0.042$  &  $ 0.011$  &  $-0.003$  &  $-0.013$  &  $ 0.066$  \\ 
	\hline
  \end{tabular}
  \end{table}